\documentclass[prd,reprint,a4paper,showpacs,superscriptaddress,11pt,onecolumn,nofootinbib]{revtex4-1}
\usepackage{setspace}
\usepackage[utf8]{inputenc}
\usepackage{bm}
\usepackage{graphics}
\usepackage{amsmath, amssymb, amsfonts}
\usepackage{natbib}
\usepackage{hyperref}
\usepackage{color}
\hypersetup{colorlinks=true}

\usepackage{graphicx}
\usepackage{sidecap}
\usepackage{subfigure}
\usepackage[T1]{fontenc}
\usepackage{stackrel}
\usepackage{float}
\usepackage{dsfont}

\newcommand{\be}{\begin{equation}}
\newcommand{\ee}{\end{equation}}
\newcommand{\bse}{\begin{subequations}}
\newcommand{\ese}{\end{subequations}}
\newcommand{\ba}{\begin{eqnarray}}
\newcommand{\ea}{\end{eqnarray}}

\newcommand{\bea}{\begin{eqnarray}}
\newcommand{\eea}{\end{eqnarray}}

\usepackage{array}
\usepackage{lineno}
\usepackage{hyperref}
\usepackage{cleveref}
\newcolumntype{L}[1]{>{\raggedright\let\newline\\\arraybackslash\hspace{0pt}}m{#1}}
\newcolumntype{C}[1]{>{\centering\let\newline\\\arraybackslash\hspace{0pt}}m{#1}}
\newcolumntype{R}[1]{>{\raggedleft\let\newline\\\arraybackslash\hspace{0pt}}m{#1}}
\begin{document}
\title{Illustrated study of the semi-holographic non-perturbative framework}

\author{Souvik Banerjee}
\email{souvik.banerjee@physics.uu.se}
\affiliation{Department of Physics and Astronomy, Uppsala University, SE-751 08 Uppsala, Sweden}
\author{Nava Gaddam}
\email{gaddam@uu.nl}
\affiliation{Institute for Theoretical Physics and Center for Extreme Matter and Emergent Phenomena, Princetonplein 5, Utrecht University, 3584 CC Utrecht, The Netherlands}
\author{Ayan Mukhopadhyay}
\email{ayan@hep.itp.tuwien.ac.at}
\affiliation{Institut f\"{u}r Theoretische Physik, Technische Universit\"{a}t Wien, Wiedner Hauptstr.~8-10, A-1040 Vienna, Austria}
\affiliation{CERN, Theoretical Physics Department, 1211 Geneva 23, Switzerland}

\begin{abstract}
Semi-holography has been proposed as an effective nonperturbative framework which can combine perturbative and nonperturbative effects consistently for theories like QCD. It is postulated that the strongly coupled nonperturbative sector has a holographic dual in the form of a classical gravity theory in the large N limit, and the perturbative fields determine the gravitational boundary conditions. In this work, we pursue a fundamental derivation of this framework particularly showing how perturbative physics by itself can determine the holographic dual of the infrared, and also the interactions between the perturbative and the holographic sectors. We firstly demonstrate that the interactions between the two sectors can be constrained through the existence of a conserved local energy-momentum tensor for the full system up to hard-soft coupling constants. As an illustration, we set up a bi-holographic toy theory where both the UV and IR sectors are strongly coupled and holographic with distinct classical gravity duals. In this construction, the requirement that an appropriate gluing can cure the singularities (geodetic incompletenesses) of the respective geometries leads us to \textit{determine} the parameters of the IR theory and the hard-soft couplings \textit{in terms of those of the UV theory}. The high energy scale behaviour of the hard-soft couplings is state-independent but their runnings turn out to be state-dependent. We discuss how our approach can be adapted to the construction of the semi-holographic framework for QCD.
\end{abstract}

\maketitle

\begin{spacing}{1.8}
\tableofcontents
\end{spacing}
\section{Introduction}

Semi-holography has been recently proposed as an effective framework in which one can include both perturbative and non-perturbative effects consistently in a wide range of energy scales. It's present formulation is targeted towards an effective description of asymptotically free theories like QCD which are weakly coupled in the ultraviolet but strongly interacting in the infrared. It is assumed that in the large $N$ limit (i) the infrared non-perturbative effects such as confinement and chiral symmetry breaking can be obtained from a holographic dual description in the form of an appropriate classical theory of gravity\footnote{This was demonstrated in the Witten-Sakai-Sugimoto top-down model \cite{Witten:1998zw,Sakai:2004cn,Sakai:2005yt} obtained from string theory. We do not assume here that the holographic description of the non-perturbative sector can be embedded in string theory.}, and (ii) the perturbative degrees of freedom determine the effective background metric, relevant and marginal couplings, and background gauge-fields (coupling to conserved currents) in which the emergent infrared holographic degrees of freedom live. The second assertion then implies that the perturbative degrees of freedom determine the leading asymptotic behaviour of the classical gravity fields forming the holographic dual of the non-perturbative sector. As we will argue, such a set-up allows for only a few effective parameters in a wide range of energy scales. Concrete phenomenological semi-holographic models with a small number of effective parameters have been proposed  for some non-Fermi liquid systems \cite{Faulkner:2010tq,Gursoy:2011gz,Gursoy:2012ie,Mukhopadhyay:2013dqa,Jacobs:2014lha,Jacobs:2015fiv}, and for the quark-gluon plasma (QGP) formed in heavy-ion collisions \cite{Iancu:2014ava,Mukhopadhyay:2015smb}. In such instances, indeed both perturbative and non-perturbative effects are phenomenologically relevant.

In this article, we will take first steps towards a derivation of the general semi-holographic framework from first principles, i.e. from the fundamental theory describing the microscopic dynamics. This amounts to answering the following questions:
\begin{itemize}
\item Which principles tell us how the perturbative degrees of freedom determine the leading asymptotic behaviours of the gravitational fields forming the holographic dual of the non-perturbative sector?
\item How do we find the appropriate classical gravity theory which provides the dual holographic description of the non-perturbative sector? 
\end{itemize}
We will arrive at partial answers to both these questions, and also illustrate the full construction of semi-holography with a toy example.

Previously, semi-holography has been conceived of as an effective simplified method for solving low energy holographic dynamics where the asymptotic geometry determining model-dependent features is replaced by simple boundary dynamics which couples to the near-horizon geometry controlling universal scaling exponents \cite{Faulkner:2010tq,Faulkner:2010jy}. It has also been argued that decoding holography as a form of non-Wilsonian RG flow which preserves Ward identities for single-trace operators (like the energy-momentum tensor) and can self-determine microscopic data via appropriate infrared endpoint conditions naturally gives rise to a more general semi-holographic framework in which the ultraviolet can be asymptotically free so that it is described by perturbative quantum field dynamics rather than by a classical gravity theory \cite{Behr:2015yna,Behr:2015aat,Mukhopadhyay:2016fre}. In this article, we will deal with the fundamental aspects of construction of the general semi-holographic framework (which may not be embeddable in string theory as we know of it today)  by understanding what constrains it structurally and illustrate it with a toy example.

The organisation of the paper is as follows. In Section \ref{sec:democracy}, we will review the present formulation of semi-holography and then argue for the need for generalising it in order for it to be an effective theory in a wide range of energy scales. In particular, we will advocate that we need a more democratic formulation where we do not give precedence to either the perturbative or to the non-perturbative degrees of freedom. Although we will call the perturbative sector as the ultraviolet sector and non-perturbative sector as the infrared sector, it is to be noted that non-perturbative effects are present even at high energy scales although these are suppressed. In principle, both sectors contribute at any energy scale although one of the sectors may give dominant contributions at a specific energy scale. Since semi-holography is a framework that is constructed at intermediate energy scales, it better treats both the ultraviolet and infrared sectors, or rather the perturbative and non-perturbative sectors in a democratic manner. Eventually the parameters of the non-perturbative  sector should be determined (perhaps not always uniquely) by the perturbative sector or vice versa. We will argue this democratic formulation is actually necessary since otherwise we cannot perform non-perturbative renormalisation of the effective parameters. We will also sketch how the democratic formulation should work.

In Section \ref{sec:couplingCFTs}, we will show how the requirement that there exists a local and conserved energy-momentum tensor constrains the effective parameters and semi-holographic coupling between the perturbative and the non-perturbative sectors. Thus we will realise a concrete democratic formulation of semi-holography at arbitrary energy scales.

In Section \ref{sec:holRGapplication}, we will illustrate the construction of semi-holography with a bi-holographic toy model in which the perturbative UV dynamics of semi-holography will be replaced by a strongly coupled holographic theory that admits a classical gravity description on its own. The infrared sector will be even more strongly coupled and also holographic. We will explicitly demonstrate the following features.
\begin{itemize}
\item Some simple consistency conditions can determine the hard-soft couplings between the two sectors and the parameters of the IR theory as functions of the parameters of the UV theory.
\item The behaviour of the hard-soft couplings in the limit $\Lambda \rightarrow \infty$ is state-independent and can be obtained from the construction of the vacuum state. However, the running of the hard-soft couplings with the scale is state-dependent.
\item The parameters defining the holographic IR classical gravity theory is fixed once and for all through the construction of the vacuum state of the full theory. However, the gravitational fields of this IR classical gravity theory undergo state-dependent field redefinitions in excited states.
\item The UV and IR classical gravity theories are both sick in the sense that the respective geometries have edge singularities (not naked curvature singularities though) arising from geodetic incompleteness. The possibility of smooth gluing of their respective edges that removes the singularities in both plays a major role in determining the full theory.
\end{itemize}
We will also examine how we can define RG flow in the bi-holographic theory.

In Section \ref{sec:outlook}, we will indicate how the steps of the construction of the bi-holographic toy theory can be applied also to the construction of the semi-holographic framework for QCD and also discuss the complications involved. Finally, we will conclude with discussions on the potential phenomenological applications of the bi-holographic framework.

\section{Democratising semiholography}\label{sec:democracy}
\subsection{A brief review}\label{review}
Let us begin by sketching a first construction of a semi-holographic model for pure large $N$ QCD based on a similar model \cite{Iancu:2014ava,Mukhopadhyay:2015smb,Mukhopadhyay:2016fkl} for the quark-gluon plasma (QGP) formed in heavy-ion collisions. The effective action for pure large $N$ QCD at a scale $\Lambda$ can be proposed to be:
\begin{align}\label{semi-hol-action}
S^{\rm QCD}[A_\mu^a, \Lambda] ~ &= ~ S^{\rm pQCD}[A_\mu^a, \Lambda] \nonumber \\
&\qquad + W^{\rm hQCD}\Big[\tilde{g}_{\mu\nu}[A_\mu^a, \Lambda], \delta\tilde{g}_{\rm YM}[A_\mu^a, \Lambda], \tilde{\theta}[A_\mu^a, \Lambda]\Big] \, ,
\end{align}
where the exact Wilsonian effective action of QCD denoted as $S^{\rm QCD}$ at a scale $\Lambda$ is composed of two parts: (i) the perturbative QCD effective action $S^{\rm pQCD}$ at the scale $\Lambda$ obtained from Feynman diagrams, and (ii) non-perturbative terms (leading to confinement) which cannot be obtained from Feynman diagrams but can be described by an emergent holographic strongly coupled QCD-like theory. The latter part of the full action is then given by $W^{\rm hQCD}$, the generating functional of the connected correlation functions of the emergent strongly coupled holographic QCD-like theory whose marginal couplings\textemdash namely $\tilde{g}_{\rm YM}$ and $\tilde{\theta}$ (or rather, their expansions around infinity and zero respectively) and the effective background metric $\tilde{g}_{\mu\nu}$ in which it lives are functionals of the perturbative gauge fields $A_\mu^a$ and the scale $\Lambda$. In order that a holographic theory can capture non-perturbative effects at even high energy scales, it must have a large number of fields as we will discuss in Section \ref{sec:outlook}. Nevertheless, at high energy scales the non-perturbative contributions are insignificant. We will argue that the semi-holographic construction can be useful at intermediate energy scales where the non-perturbative effects can also be captured by a few gravitational field via holography to a good degree of approximation.

It is to be noted that $W^{\rm hQCD}$ should be defined with an appropriate vacuum subtraction so that it vanishes when the modifications in the couplings $\delta\tilde{g}_{\rm YM}$ and $\tilde{\theta}$ vanish, and when\footnote{More generally, the subtraction should ensure that $W^{\rm hQCD}$ vanishes when $\tilde{g}_{\mu\nu}$ is identical to the fixed background metric where all the degrees of freedom live.} $\tilde{g}_{\mu\nu} = \eta_{\mu\nu}$. Asymptotically when $\Lambda \rightarrow \infty$, the sources for the emergent holographic QCD are also expected to vanish, so that the full action receives perturbative contributions almost exclusively. In the infra-red, however, the holographic contributions are expected to dominate. 

In the large $N$ limit, the emergent holographic QCD is expected to be described by a classical gravitational theory. Therefore,
\begin{align}\label{on-shell-grav}
W^{\rm hQCD}\Big[\tilde{g}_{\mu\nu}[A_\mu^a], \delta\tilde{g}_{\rm YM}[A_\mu^a], \tilde{\theta}_{\rm YM}[A_\mu^a]\Big] ~ &= ~ S^{\rm on-shell}_{grav}\left[\tilde{g}_{\mu\nu}= g^{\rm (b)}_{\mu\nu},\right. \nonumber \\
&\qquad\qquad\left. \delta\tilde{g}_{\rm YM} = \phi^{(b)}, \tilde{\theta} = \chi^{\rm (b)}\right] \, ,
\end{align}
i.e. $W^{\rm hQCD}$ is to be identified with the on-shell action $S^{\rm on-shell}_{grav}$ of an appropriate five-dimensional classical gravity theory consisting of \textit{at least} three fields, namely the metric $G_{MN}$, the dilaton $\Phi$ and the axion $\mathcal{X}$. Furthermore, the leading behaviour of the bulk metric is given by its identification with the boundary metric $g^{\rm (b)}_{\mu\nu}$, whilst $\delta\tilde{g}_{\rm YM}$ is identified with the boundary value $\phi^{\rm (b)}$ of the bulk dilaton $\Phi$ and $\tilde{\theta}$ is identified with the boundary value $\chi^{\rm (b)}$ of the bulk axion $\mathcal{X}$. For reasons that will soon be elucidated, one may now postulate that:
\begin{subequations}
\label{allequations}
\begin{eqnarray}
g^{\rm (b)}_{\mu\nu} ~ &=& ~ \eta_{\mu\nu} + \gamma t^{\rm pQCD}_{\mu\nu}, ~~ {\rm with} ~~~~ t^{\rm pQCD}_{\mu\nu} ~ = ~  \frac{2}{\sqrt{-g}}\frac{\delta S^{\rm pQCD}[A_\mu^a, \Lambda]}{\delta g_{\mu\nu}}\left.\right|_{g_{\mu\nu} = \eta_{\mu\nu}} \label{source-1} \\
\phi^{\rm (b)} ~ &=& ~ \beta h^{\rm pQCD}, \qquad\quad {\rm with} ~~~~ h^{\rm pQCD} ~ = ~ \frac{\delta S^{\rm pQCD}[A_\mu^a, \Lambda]}{\delta g^{\rm YM}[\Lambda]} \, , \quad \label{source-2}\\
\chi^{\rm (b)} ~ &= &~ \alpha a^{\rm pQCD}, \qquad\quad {\rm with} ~~~~ a^{\rm pQCD} ~ = ~ \frac{\delta S^{\rm pQCD}[A_\mu^a, \Lambda]}{\delta \theta}\label{source-3} \, . 
\end{eqnarray}
\end{subequations}
The couplings $\alpha$, $\beta$ and $\gamma$ have been called hard-soft couplings. These of course cannot be new independent parameters, but rather of the functional forms $(1/\Lambda^4) f(\Lambda_{\rm QCD}/\Lambda)$ which should be derived from first principles. Furthermore, $f(0)$ must be finite so that the non-perturbative contributions to the full action vanish in the limit $\Lambda \rightarrow \infty$ reproducing asymptotic freedom. If $S^{\rm pQCD}$ were just the classical Yang-Mills action, then \cite{Mukhopadhyay:2015smb}:
\begin{align}\label{t-cl}
t^{\rm pQCD}_{\mu\nu} ~ &=  ~ \frac{1}{N_c} {\rm tr} \left(F_{\mu\alpha}F_\nu^{\phantom{\mu}\alpha} - \frac{1}{4}\eta_{\mu\nu}F_{\alpha\beta}F^{\alpha\beta}\right),\nonumber\\
h^{\rm pQCD} ~ &= ~ \frac{1}{4N_c} {\rm tr}\left(F_{\alpha\beta}F^{\alpha\beta}\right),\nonumber\\
a^{\rm pQCD} ~ &= ~ \frac{1}{4N_c} {\rm tr}\left(F_{\alpha\beta}\tilde{F}^{\alpha\beta}\right) \, .
\end{align}
It can be readily shown that in consistency with the variational principle, the full semi-holographic action \eqref{semi-hol-action} can be written in the following form:
\begin{equation}\label{modified-action}
S[A_\mu^a, \Lambda] = S^{\rm pQCD}[A_\mu^a, \Lambda] +\frac{1}{2} \int {\rm d}^4x \, \overline{\mathcal{T}}^{\mu\nu}g^{(b)}_{\mu\nu} + \int {\rm d}^4x \, \overline{\mathcal{H}}\phi^{\rm (b)} + \int {\rm d}^4x \, \overline{\mathcal{A}}\chi^{(b)} \, ,
\end{equation}
where 
\begin{eqnarray}\label{hol-ops}
\overline{\mathcal{T}}^{\mu\nu} = 2 \frac{\delta S^{\rm on-shell}_{grav}}{\delta g^{\rm (b)}_{\mu\nu}}, \quad \overline{\mathcal{H}} = \frac{\delta S^{\rm on-shell}_{grav}}{\delta \phi^{\rm (b)}}, \quad \overline{\mathcal{A}} = \frac{\delta S^{\rm on-shell}_{grav}}{\delta \chi^{\rm (b)}},
\end{eqnarray}
are the self-consistent expectation values of the holographic operators which are non-linear and non-local functionals of the sources\footnote{It is assumed here that the full theory lives in flat Minkowski space with metric $\eta_{\mu\nu}$. It is easy to generalise the construction to any metric on which the full degrees of freedom live. For details, see \cite{Mukhopadhyay:2015smb}.} $g^{\rm (b)}_{\mu\nu}$, $\phi^{\rm (b)}$ and $\chi^{\rm (b)}$. 

The reason for postulating the hard-soft interactions to be of the forms given by equations \eqref{source-1} to \eqref{source-3} can now be readily explained. We need to solve for the full dynamics in an iterative fashion (assuming that the iteration indeed converges). This means that the dynamics of the perturbative sector is modified by the holographic operators which appear as self-consistent mean fields as in \eqref{modified-action}. The holographic operators are in turn obtained by solving classical gravity equations with sources given by \eqref{source-1}, \eqref{source-2} and \eqref{source-3}, which are determined by the perturbative gauge fields. It should therefore be guaranteed that both sectors must be solvable at each step in the iteration including perturbative quantum effects. Therefore, both the perturbative and non-perturbative sectors should be renormalizable at each step of the iteration so that one can solve for the dynamics of both without introducing any new coupling. The modified perturbative action (\ref{modified-action}) in the limit $\Lambda \rightarrow \infty$ is indeed a \textit{marginal} deformation of the standard perturbative QCD action since the added terms involve $t^{\rm cl}_{\mu\nu}$, $h^{\rm cl}$ and $a^{\rm cl}$ which are all possible (scalar and tensor) gauge-invariant operators of mass dimension four. Furthermore, this is also why the gravitational theory contains sources for only the dimension four operators as in \eqref{on-shell-grav}, despite the possible existence of many other (massive) gravitational fields generating (non-perturbative) condensates of higher dimensional operators without additional sources. 

Finally, it is important to reiterate that the importance of the hard-soft couplings given in \eqref{source-1}, \eqref{source-2} and \eqref{source-3} relies on the emergence of an intermediate scale $\Lambda_{\rm I} > \Lambda_{\rm QCD}$ between the energy scales where we can rely exclusively on either perturbative QCD or chiral Lagrangian effective field theories which can be reproduced from holographic models such as \cite{Witten:1998zw,Sakai:2004cn,Sakai:2005yt}. This intermediate scale $\Lambda_{\rm I}$ is most likely where the perturbative gauge coupling is of order unity but not too large. At this intermediate scale itself, the hard-soft couplings should give significant modifications to perturbative dynamics.

In the context of an application to QGP, $S^{\rm pQCD}$ can be replaced by the glasma effective action, i.e. a classical Yang-Mills action for the small-$x$ saturated gluons (which form a weakly coupled over-occupied system) with colour sources provided by the large-$x$\footnote{Here $x$ denotes the fraction of hadronic longitudinal momentum carried by the partonic gluon.} ($x> x_0$) gluons (frozen on the time-scale of the collisions) with a distribution function whose evolution with the cut-off $x_0$ can be followed perturbatively \cite{Iancu:2000hn,Lappi:2006fp,Gelis:2010nm}. In this case, one can show that (i) the full system has a well-defined local energy momentum tensor that is conserved in flat space, and (ii) the iterative method of solving the full dynamics indeed converges at least in some simple test cases \cite{Mukhopadhyay:2015smb}. 

\subsection{Why and how should we democratise semi-holography?}\label{whyandhow}

The present formulation of semi-holography as discussed above makes two central assumptions which are:
\begin{enumerate}
\item The perturbative sector and the holographic theory dual to the non-perturbative sector are both deformed marginally with scalar and tensorial couplings which are functionals of the operators of the other sector.
\item The full action (as for instance \eqref{semi-hol-action} in the case of pure large-$N$ QCD) can be written as a functional of the perturbative fields only.\footnote{Note that $W^{\rm hQCD}$ in \eqref{semi-hol-action} is after all a functional of the sources alone which in turn are functionals of the perturbative gauge fields. Furthermore, the term \textit{perturbative fields} may also seem problematic because after their coupling with the non-perturbative sector, these do not remain strictly perturbative. Nevertheless, if we solve the two sectors via iteration as discussed before, we can treat the dynamics of these fields perturbatively (with self-consistently modified couplings) at each step of the iteration as discussed before.}
\end{enumerate}
In what follows, we will argue that if semi-holography needs to work in the sense of a non-perturbative effective framework, we must replace the second assumption with the simple assertion that:
\begin{itemize}
\item The full theory must have a conserved and local energy-momentum tensor which can be constructed without the need to know the explicit Lagrangian descriptions of the effective ultraviolet or infrared dynamics.
\end{itemize}
It is to be noted that the formulation discussed above already leads to a local and conserved energy-momentum tensor of the combined system which can be constructed explicitly. As shown in \cite{Mukhopadhyay:2015smb}, this can be obtained from the complete action \eqref{semi-hol-action} by differentiating it with respect to the fixed background $g_{\mu\nu}$ on which the full system lives. One can prove that this energy-momentum tensor is conserved in the fixed background $g_{\mu\nu}$ when the full system is solved, i.e when the Ward identity of the CFT in the dynamical background $\tilde{g}_{\mu\nu}$ and the modified dynamical equations of the perturbative sector are satisfied simultaneously. \textit{The crucial point is that we must not insist that we can derive the full energy-momentum tensor from an action such as \eqref{semi-hol-action} which is a functional of the ultraviolet fields only. We will argue that such a demand automatically arises from a democratic formulation of semi-holography where the full energy-momentum tensor can be constructed directly from the Ward identities of the perturbative and non-perturbative sectors in respective background metrics and sources that are determined by the operators of the other sector. This full energy-momentum tensor should be conserved locally in the fixed background metric where the full system lives.}

We lay out our reasons for advocating democratic formulation of semi-holography. The first one is somewhat philosophical. In the case we are assuming that the strongly coupled non-perturbative sector is described by a holographically dual classical gravity theory, we are not making any explicit assumption of the Lagrangian description of this dual holographic theory. It could be so in certain situations we do not know the explicit Lagrangian description of the perturbative sector as well.\footnote{Note that by assuming that we have a weakly coupled perturbative description we do not necessarily commit ourselves to a Lagrangian description also. Such a perturbation series can be obtained by a chain of dualities without a known Lagrangian description as in the case of some quiver gauge theories \cite{Alday:2009aq}.} We should be able to formulate an explicit semi-holographic construction in such a situation also. Furthermore, the basic idea of semi-holography is to take advantage of dualities. It is quite possible that the ultraviolet is strongly coupled instead of the infrared and we should take advantage of a weakly coupled (perhaps holographic) dual description of the ultraviolet. In that case, the original Lagrangian description of the UV even if known will not be useful. Therefore, we better have a broader construction which can work without the need of knowing an explicit Lagrangian description of the perturbative sector or that of the holographic theory dual to the non-perturbative sector. This implies we need to treat both on equal footing.

The second reason for advocating democratic formulation is more fundamental. Let us take the example where in the infrared the theory flows to a strongly coupled holographic conformal field theory (IR-CFT) from a weakly coupled fixed point (UV-CFT) in the ultraviolet. There will be a specific UV-IR operator map in such a theory relating operators in the UV fixed point to those defined at the IR fixed point via scale-evolution as explicitly known in the case of RG flows between minimal-model two-dimensional conformal field theories \cite{Zamolodchikov:1987ti,Zamolodchikov:1987jf,Ludwig:1987gs}. Typically a relevant operator in the UV will flow to an irrelevant operator in the IR. In fact the entire flow will be generated by a relevant deformation of the UV fixed point -- the operator(s) generating such a deformation will become irrelevant at the IR fixed point. Let the UV deformation be due to a coupling constant $g$ multiplying a relevant operator $O^{\rm UV}$ which will flow to an irrelevant operator $O^{\rm IR}$ of the strongly coupled holographic IR-CFT that is represented by a bulk field $\Phi$. A naive way to formulate semi-holography in such a case will be:
\begin{equation}\label{example}
S = S^{\rm UV-CFT} + g \int{\rm d}^4x \, O^{\rm UV} + S^{\rm grav}[\phi^{\rm (b)} = \lambda O^{\rm UV} ],
\end{equation} 
where $\phi^{\rm (b)}$ is the coefficient of the leading asymptotic term of the bulk field $\Phi$, $\lambda$ is a dimensionful hard-soft coupling constant and we have suppressed the scale dependence. The immediate problem is that we have turned on an irrelevant deformation of the dual holographic theory as
\begin{equation}
S^{\rm grav}[\phi^{\rm (b)} = \lambda O^{\rm UV} ] =\lambda \int {\rm d}^4 x\, O^{\rm UV}O^{\rm IR}
\end{equation}
represents an irrelevant deformation of the IR-CFT since $O^{\rm IR}$ is an irrelevant operator. This contradicts our assumption that both sectors should be renormalizable after being mutually deformed by the operators of the other. Furthermore, turning on non-trivial sources of irrelevant operators leads to naked gravitational singularities in holography and the removal of asymptotic anti-de Sitter behaviour of the spacetime. We can therefore argue that the formulation of the full semi-holographic theory in terms of an action which is a functional of the UV variables as in \eqref{example} should be abandoned as it leads to such a contradiction. As we will show later in the democratic formulation, we will be able to generate a non-trivial expectation value $\langle O^{\rm IR}\rangle$ without sourcing it.

Similarly, in the case of QCD where the infrared dynamics has a mass gap, non-perturbative vacuum condensates of operators of high mass dimensions are crucial for the cancellation of renormalon Borel poles of perturbation series \cite{Parisi:1978bj,Shifman:1978bx,citeulike:9712091}. This implies that we need to couple irrelevant operators of the infrared holographic theory with the gauge-invariant marginal operators of perturbative QCD as we will discuss in Section \ref{sec:outlook}. This is not quite possible within the present formulation for the same reasons mentioned above.

Let us then sketch how the democratic formulation should be set-up. Let us denote $S^{(1)}$ as the quantum effective perturbative action and $S^{(2)}$ as the quantum effective action of the holographic theory dual to the non-perturbative sector both defined at the same energy scale $\Lambda$. Furthermore, for simplicity let us assume that the two sectors couple via their energy-momentum tensors and a scalar operator in each sector. The democratic formulation postulates that the individual actions are deformed as follows:\footnote{Note that we have not turned on sources for the elementary fields. Therefore, $S^{(i)}$ can denote either $W$, the generating functional of the connected correlation functions, or $\Gamma$, the 1-PI (one-particle irreducible) effective action, which is the Legendre transform of $W$. In absence of sources for elementary fields, $W = \Gamma$. Below, the effective actions of both theories have been defined in specific background metrics and with specific couplings (denoted as $J^{(i)}$ and corresponding to specific composite operator vertices) which are functionals of the operators of the two sectors. These effective actions can be defined even when the Lagrangian descriptions (i.e. representations of the two sectors via some elementary fields) are unknown.}
\begin{eqnarray}\label{coupling-gen-1}
S^{(1)} = S^{(1)}[g^{(1)}_{\mu\nu}, J^{(1)}], \quad S^{(2)} = S^{(2)}[g^{(2)}_{\mu\nu}, J^{(2)}]
\end{eqnarray}
with 
\begin{eqnarray}\label{coupling-gen-2}
g^{(i)}_{\mu\nu} &=& \eta_{\mu\nu} + h^{(i)}_{\mu\nu}[t^{(i)}_{\mu\nu}, O^{(i)}], \nonumber\\
J^{(i)} &=& F^{(i)}[t^{(i)}_{\mu\nu}, O^{(i)}],
\end{eqnarray}
where $i$ denotes $1,2$. This means that the two sectors couple only via their effective sources and background metrics. Furthermore \textit{we should not allow redundant dependencies} meaning that $h^{(1)}_{\mu\nu}$ and $F^{(1)}$ can depend on $t^{(1)}_{\mu\nu}$ and $O^{(1)}$ only such that when $t^{(2)}_{\mu\nu} = O^{(2)} = 0$, then it should also follow that $h^{(1)}_{\mu\nu} = F^{(1)} = 0$. The aims will be:
\begin{enumerate}
\item To determine the functional forms of $h^{(i)}_{\mu\nu}$ and $F^{(i)}$ by requiring the existence of a local energy-momentum tensor of the full theory conserved in the background metric $\eta_{\mu\nu}$ where the full theory lives and disallowing redundant dependencies, and
\item To determine the theory $S^{(2)}$ and the hard-soft coupling constants appearing in $h^{(i)}_{\mu\nu}$ and $F^{(i)}$ as functions of the parameters of the perturbative sector, i.e. the parameters in $S^{(1)}$.
\end{enumerate}
Remarkably, we will see in the following section that the requirement of the existence of a local and conserved energy-momentum tensor of the full system along with some other simple assumptions constrains the functional forms of $h^{(i)}_{\mu\nu}$ and $F^{(i)}$ such that we can only have a few possible hard-soft coupling constants relevant for physics (including non-perturbative effects) at given energy scales. The scale ($\Lambda-$)dependence of $h^{(i)}_{\mu\nu}$ and $F^{(i)}$ should be only through the hard-soft coupling constants. \textit{If the operator $O^{(2)}$ in the holographic theory dual to the infrared is an irrelevant operator then we should demand $J^{(2)} = 0$. The functional form of $F^{(2)}$ will then play a major role in determining how the hard-soft coupling constants and parameters of the holographic theory (i.e. parameters in $S^{(2)}$) are determined by the parameters in $S^{(1)}$.} In order to demonstrate how we can achieve the second task of determining the hard-soft coupling constants and the parameters in $S^{(2)}$ in principle, we will construct a toy model in Section \ref{sec:holRGapplication}. Later in Section \ref{sec:outlook} we will outline how we can achieve this in the case of QCD. This of course will be a difficult problem in practice, and therefore we will postpone this to the future.

As will be clear in the next section, even if we can choose an arbitrary background metric $g_{\mu\nu}$ instead of $\eta_{\mu\nu}$ for the full system, we can construct the combined local and conserved energy-momentum tensor. Furthermore, it will be trivial to generalise the construction to the case where there are multiple relevant/marginal scalar operators in the perturbative sector. We will not consider the case when the perturbative sector has relevant/marginal vector operators and tensor operators other than the energy-momentum tensor. We will postpone such a study to the future. The phenomenological semi-holographic constructions discussed in the previous subsection will turn out to be special cases of the more general scenario to be described below.

\section{Coupling the hard and soft sectors}\label{sec:couplingCFTs}
In what follows, we will study how we can determine the most general form of couplings between the hard and soft sectors following \eqref{coupling-gen-1} and \eqref{coupling-gen-2} such that there exists a local and conserved energy-momentum tensor of the full system in the fixed background metric. We will not assume any Lagrangian description of either sector in terms of elementary quantum fields and therefore we will only use the local Ward identities of each sector. Furthermore, we will disallow redundant dependencies in the coupling functions which simply redefine the respective effective UV and IR theories as discussed above.

For the moment, we will assume that the UV theory (perturbative sector) has one relevant operator $O^{(1)}$ which couples to a relevant/marginal/irrelevant operator $O^{(2)}$ in the IR theory (the holographic non-perturbative sector).  Furthermore, we should also take into account the energy-momentum tensor operators $T^{(1)\mu\nu}$ and $T^{(2)\mu\nu}$ in the coupling of the UV and IR theories. As mentioned earlier, it will be clear later how we can generalise our results to the case of multiple relevant and marginal scalar operators in the perturbative sector each coupling to multiple operators in the non-perturbative sector.

\subsection{Simple scalar couplings}
The simplest possible consistent coupling of the UV and IR theories leading to a conserved local energy-momentum tensor of the full system is given by:
\begin{equation}\label{eqn:teasercouplings}
g^{(1)}_{\mu\nu}  ~ = ~ g^{(2)}_{\mu\nu}  ~ = ~ g_{\mu\nu}, \quad J^{(1)} ~ = ~ \alpha_0 O^{(2)}, \quad J^{(2)} ~ = ~ \alpha_0 O^{(1)}.
\end{equation}
Above $\alpha_0$ is once again a scale-dependent dimensionful hard-soft coupling constant. In the case we have a fixed point both in the UV and in the IR, we can postulate
\begin{equation}\label{scale-dependence}
\alpha = A_0 \frac{1}{\Lambda_{\rm I}^{\Delta^{\rm UV} + \Delta^{\rm IR}- d}} + \frac{1}{\Lambda^{\Delta^{\rm UV} + \Delta^{\rm IR}- d}}\,\, f\left(\frac{\Lambda}{\Lambda_{\rm I}}\right),
\end{equation}
where $A_0$ is a dimensionless constant, $\Delta^{\rm UV}$ is the scaling dimension of the UV operator $O^{(1)}$ at the UV fixed point, $\Delta^{\rm IR}$ is the scaling dimension of the IR operator $O^{(2)}$ at the IR fixed point and $d$ is number of spacetime dimensions. Furthermore, $\Lambda_{\rm I}$ is an emergent intermediate energy-scale (but different from $\Lambda_{\rm QCD}$ in case of QCD). If the UV and/or IR limits are not conformal, then the $\Lambda-$dependence of $\alpha_0$ should be more complicated. At present we will not bother about $\Lambda-$dependence of the hard-soft couplings although we should keep in mind that the effective $\Lambda-$dependence of the couplings of the two sectors arises via them.

Both the UV and IR theories will have their respective Ward identities:
\begin{equation}\label{WIs}
\nabla_\mu T^{(1)\mu}_{\phantom{(1)\mu}\nu} ~ = ~ O^{(1)}\nabla_\nu J^{(1)} \qquad \text{and} \qquad \nabla_\mu T^{(2)\mu}_{\phantom{(2)\mu}\nu} ~ = ~ O^{(2)}\nabla_\nu J^{(2)},
\end{equation}
where $\nabla$ is the covariant derivative constructed in the fixed background metric $g_{\mu\nu}$, $T^{(1)\mu}_{\phantom{(1)\mu}\nu}= T^{(1)\mu\rho}g_{\rho\nu}$, etc. These identities will be satisfied once we have solved the full dynamics self-consistently. 

It is clear that these Ward identities together with $J^{(1)}$ and $J^{(2)}$ specified via \eqref{eqn:teasercouplings} imply the existence of $T^{\mu}_{\phantom{\mu}\nu}$ defined as
\begin{equation}\label{fullTsimple}
T^\mu_{\phantom{\mu}\nu} ~ := ~ T^{(1)\mu}_{\phantom{(1)\mu}\nu} + T^{(2)\mu}_{\phantom{(2)\mu}\nu} - \alpha O^{(1)}O^{(2)} \delta^\mu_{\phantom{\mu}\nu} 
\end{equation}
which satisfies the combined Ward identity
\begin{equation}\label{eqn:teaserwardidentity}
\nabla_\mu T^{\mu}_{\phantom{\mu}\nu} = 0.
\end{equation}
Therefore, $  T^{\mu\nu} =T^{\mu}_{\phantom{\mu}\rho}g^{\rho\nu}$ can be identified with the local conserved energy-momentum tensor of the full system. \textit{The crucial point is that the forms of the sources specified via \eqref{eqn:teasercouplings} imply that the respective Ward identities \eqref{WIs} add to form a total derivative and thus results in a conserved energy-momentum tensor \eqref{fullTsimple} for the full system.}

Of course, we can make other choices for  \eqref{eqn:teasercouplings}. One example is
\begin{equation}\label{eqn:teasercouplings2}
g^{(1)}_{\mu\nu}  ~ = ~ g^{(2)}_{\mu\nu}  ~ = ~ g_{\mu\nu}, \quad J^{(1)} ~ = ~ \alpha O^{(2)} + \frac{1}{2}\tilde\alpha_1 O^{(1)}, \quad J^{(2)} ~ = ~ \alpha O^{(1)} + \frac{1}{2}\tilde\alpha_2 O^{(2)}.
\end{equation}
In this case, we would have obtained
\begin{equation}\label{fullTsimple2}
T^\mu_{\phantom{\mu}\nu} ~ = ~ T^{(1)\mu}_{\phantom{(1)\mu}\nu} + T^{(2)\mu}_{\phantom{(2)\mu}\nu} -\left( \alpha O^{(1)}O^{(2)} -\tilde{\alpha}_1{O^{(1)}}^2 + \tilde{\alpha}_2{O^{(2)}}^2\right)  \delta^\mu_{\phantom{\mu}\nu}.
\end{equation}
This would have led to redundancies as $\tilde{\alpha}_i$ simply lead to redefinitions of the UV and IR theories. Therefore, without loss of generality we will not allow such parameters. Another possibility is:
\begin{equation}\label{eqn:teasercouplings3}
g^{(1)}_{\mu\nu}  ~ = ~ g^{(2)}_{\mu\nu}  ~ = ~ g_{\mu\nu}, \quad J^{(1)} ~ = ~ \alpha_1 {O^{(2)}}^2 O^{(1)}, \quad J^{(2)} ~ = ~ \alpha_1 {O^{(1)}}^2 O^{(2)} ,
\end{equation}
with $\alpha_1$ being an appropriate scale-dependent constant. In this case, 
\begin{equation}\label{fullTsimple3}
T^\mu_{\phantom{\mu}\nu} ~ = ~ T^{(1)\mu}_{\phantom{(1)\mu}\nu} + T^{(2)\mu}_{\phantom{(2)\mu}\nu} -\frac{3}{2}\alpha_1 \left(O^{(1)}O^{(2)}\right)^2 \delta^\mu_{\phantom{\mu}\nu}.
\end{equation}
In fact, one can more generally choose\footnote{It is easy to see that the general class of such simple scalar couplings should be such that $O^{(1)}[J^{(1)}, J^{(2)}]\, {\rm d}J^{(1)}+O^{(2)}[J^{(1)}, J^{(2)}] \,{\rm d}J^{(2)} $ should be a total differential in which we have inverted the functions $J^{(1)}[O^{(1)}, O^{(2)}] $ and $J^{(2)}[O^{(1)}, O^{(2)}] $ to obtain $O^{(1)}[J^{(1)}, J^{(2)}]$ and $O^{(2)}[J^{(1)}, J^{(2)}]$. If $J^{(1)}[O^{(1)}, O^{(2)}] $ and $J^{(2)}[O^{(1)}, O^{(2)}] $ are analytic in $O^{(1)}$ and $O^{(2)}$ at $O^{(1)} = O^{(2)} = 0$, we obtain the general expressions below.}
\begin{equation}\label{eqn:teasercouplingsgen}
g^{(1)}_{\mu\nu}  ~ = ~ g^{(2)}_{\mu\nu}  ~ = ~ g_{\mu\nu}, \quad J^{(1)} ~ = ~ \sum_{k= 0}^\infty\alpha_k {O^{(2)}}^{k+1} {O^{(1)}}^k, \quad J^{(2)} ~ = ~ \sum_{k= 0}^\infty \alpha_k {O^{(1)}}^{k+1} {O^{(2)}}^k.
\end{equation}
 and this will lead to a conserved energy-momentum tensor for the full system given by 
 \begin{equation}\label{fullTsimplegen}
T^\mu_{\phantom{\mu}\nu} ~ = ~ T^{(1)\mu}_{\phantom{(1)\mu}\nu} + T^{(2)\mu}_{\phantom{(2)\mu}\nu} -\sum_{k= 0}^\infty\frac{2k+1}{k+1}\alpha_k \left(O^{(1)}O^{(2)}\right)^{k+1} \delta^\mu_{\phantom{\mu}\nu}.
\end{equation}
satisfying:
\begin{equation}
\nabla_\mu T^{\mu}_{\phantom{\mu}\nu} = 0.
\end{equation}
Our general expectation is that in QCD, $\langle O^{(i)} \rangle \approx \Lambda_{\rm QCD}^\kappa$ whereas $\alpha_k \approx \Lambda_{\rm I}^{\kappa'}$ (for appropriate $\kappa$ and $\kappa'$), where $\Lambda_{\rm I}$ is a state-dependent scale such that $\Lambda_{\rm QCD} \ll \Lambda_{\rm I}$. In case of the vacuum, $\Lambda_{\rm I}$ could be the scale where the strong coupling is order unity (i.e. neither too small nor too large) and in case of QGP formed in heavy ion collisions $\Lambda_{\rm I}$ could be the saturation scale. If this assumption is true, one can make a useful truncation in $k$ in \eqref{eqn:teasercouplingsgen}, as the terms neglected will be suppressed by higher powers of $\Lambda_{\rm QCD}/\Lambda_{\rm I}$. In this case, semi-holography will turn out to be an useful effective non-perturbative framework. Of course the hard-soft couplings $\alpha_k$s and the condensates $\langle O^{(i)} \rangle$s are both scale-dependent. However, as long as their scale dependence do not spoil the above justification for the truncation of terms that appear in the couplings of the two sectors, semi-holography can be used as an effective framework at least for a class of processes.

It is to be noted that only in the simplest case, i.e. when $\alpha_k = 0$ for $k\neq 0$, we may be able to reproduce the full energy-momentum tensor \eqref{fullTsimplegen} and the sources \eqref{eqn:teasercouplingsgen} from an action. In this case, the action is given by:
\be
\label{Lint}
S =S^{(1)} + S^{(2)} + \alpha_0 \int d^d x \, O^{(1)}O^{(2)}.
\ee
For other cases, one can reproduce the energy-momentum tensor but cannot reproduce the right sources. However, even in the case $\alpha_k = 0$ for $k\neq 0$, the action \eqref{Lint} can reproduce the energy-momentum tensor only when $O^{(1)}$ and $O^{(2)}$ are composites of elementary scalar fields with no derivatives involved. This observation has been made earlier in \cite{Aharony:2006hz} in a different context. The lesson is that an action of the form \eqref{Lint} does not exist in the general semi-holographic formulation of non-perturbative dynamics. In fact, if such an action existed, it would have been problematic as it would have implied doing a naive path integral over both UV and IR fields. This would not have been desirable because the IR degrees of freedom are \textit{shadows} of the UV degrees of freedom in the sense that they do not have independent existence. After all, the IR theory and hard-soft couplings should be determined by the coupling constants of perturbation theory governing the UV dynamics. Since the hard-soft couplings are state-dependent, these and the parameters of the IR theory, generally speaking, should be determined by how perturbative dynamics describe the state or rather participate in the process being measured. We will examine this feature in our toy model illustration in the following section.

\subsection{More general scalar couplings}\label{sec:scalar-scalar-coupling}
So far, we have considered the cases in which the effective background metrics for the UV and IR theories are identical to the fixed background metric $g_{\mu\nu}$ in which all degrees of freedom live. Here, we will examine the cases when the effective background metrics $g^{(1)}_{\mu\nu}$ and $g^{(2)}_{\mu\nu}$ of the UV and IR theories are different and have operator-dependent scale factors. In this case, we will consider
\begin{eqnarray}\label{eqn:g1g2}
\tilde{g}^{(1)}_{\mu\nu} ~ = ~ g_{\mu\nu}e^{2\sigma^{(1)}\left[O^{(i)}, \, T^{(i)}\right]}, \quad \tilde{g}^{(2)}_{\mu\nu} ~ = ~ g_{\mu\nu}e^{2\sigma^{(2)}\left[O^{(i)}, \, T^{(i)}\right]},\nonumber\\ J^{(i)} = J^{(i)}[O^{(j)}, T^{(j)}].
\end{eqnarray}
Above, $T^{(i)} \equiv T^{(i)\mu\nu} g^{(i)}_{\mu\nu}$ is the trace of the energy-momentum tensors in the respective effective background metrics. Once again, we will disallow redundant dependencies of the sources on the operators.

Let us first establish a useful identity. The Ward identity for the local conservation of energy and momentum in the background metric $\tilde{g}_{\mu\nu} = g_{\mu\nu} e^{2\sigma}$, i.e.
\begin{equation}
\tilde{\nabla}_\mu T^{\mu}_{\phantom{\mu}\nu} = O \tilde{\nabla}_\nu J,
\end{equation}
where $\tilde\nabla$ is the covariant derivative constructed from $\tilde{g}$ can be rewritten as
\begin{equation}\label{eqn:scalarwardidentity}
\nabla_\mu \left(T^{\mu}_{\phantom{\mu}\nu}e^{d\sigma}\right) - \frac{1}{d} ({\rm Tr} \, T)\, \nabla_\nu e^{d\sigma}-e^{d\sigma}O \nabla_\nu J = 0 \, ,
\end{equation}
where $\nabla$ is built out of $g_{\mu\nu}$ and $d$ is the number of spacetime dimensions. The Ward identity in this form will be useful for the construction of the energy-momentum tensor of the full system which should be locally conserved in the background $g_{\mu\nu}$.

The general consistent scalar-type couplings which give rise to a conserved energy-momentum tensor of the full system then are of the form:
\begin{subequations}
\label{allequations}
\begin{align}
e^{d\sigma^{(1)}} ~ &= ~ 1 +d \beta \left(T^{(2)} + O^{(2)}\right) ,\label{eqn:sigma1}\\
e^{d\sigma^{(2)}} ~ &= ~ 1+ d \beta  \left(T^{(1)} + O^{(1)}\right), \label{eqn:sigma2}\\
J^{(1)} ~ &= ~ \frac{1}{d}{\rm ln} \left(1 +d\beta\left(T^{(2)} + O^{(2)}\right)\right) \nonumber\\&+ \sum_{k = 0}^\infty\alpha_k {O^{(2)}}^{k+1}\left(1 +d\beta\left(T^{(1)} + O^{(1)}\right)\right)^{k+1}  {O^{(1)}}^k \left(1 +d\beta\left(T^{(2)} + O^{(2)}\right)\right)^k , \label{eqn:J1}\\
J^{(2)} ~ &= ~ \frac{1}{d}{\rm ln} \left(1 + d\beta\left(T^{(1)} + O^{(1)}\right)\right) \nonumber\\&+ \sum_{k = 0}^\infty\alpha_k {O^{(1)}}^{k+1}\left(1 +d\beta\left(T^{(2)} + O^{(2)}\right)\right)^{k+1}  {O^{(2)}}^k \left(1 +d\beta\left(T^{(1)} + O^{(1)}\right)\right)^k\, . \label{eqn:J2}
\end{align}
\end{subequations}
Clearly when $\beta = 0$ we revert back to the case \eqref{eqn:teasercouplingsgen} discussed in the previous subsection.

Using \eqref{eqn:scalarwardidentity}, we can rewrite the Ward identities in the respective UV and IR theories in the form:
\begin{align}
\nabla_\mu \left(T^{(1)\mu}_{\phantom{(1)\mu}\nu}e^{d\sigma^{(1)}}\right) - \frac{1}{d} T^{(1)} \, \nabla_\nu e^{d\sigma^{(1)}}-e^{d\sigma^{(1)}}O^{(1)} \nabla_\nu J^{(1)} ~ &= ~ 0, \nonumber\\
\nabla_\mu \left(T^{(2)\mu}_{\phantom{(2)\mu}\nu}e^{d\sigma^{(2)}}\right) - \frac{1}{d} T^{(2)} \, \nabla_\nu e^{d\sigma^{(1)}}-e^{d\sigma^{(2)}}O^{(2)} \nabla_\nu J^{(2)} ~ &= ~ 0,
\end{align}
where $T^{(i)\mu}_{\phantom{(i)\mu}\nu} \equiv T^{(i)\mu\nu}g^{(i)}_{\mu\nu}$. Substituting \eqref{eqn:g1g2}, and then \eqref{eqn:sigma1}, \eqref{eqn:sigma2}, \eqref{eqn:J1} and \eqref{eqn:J2} in the above equations, we find that
\begin{align}
T^{\mu}_{\phantom{\mu}\nu} ~ &= ~ T^{(1)\mu}_{\phantom{(1)\mu}\nu}\left(1 + d \beta \left(T^{(2)} + O^{(2)}\right)\right) ~ + ~ T^{(2)\mu}_{\phantom{(1)\mu}\nu}\left(1 + d \beta \left(T^{(1)} + O^{(1)}\right)\right) \nonumber\\ 
&\qquad - \beta\left(T^{(1)} + O^{(1)}\right)\left(T^{(2)} + O^{(2)}\right)\delta^{\mu}_{\phantom{\mu}\nu} \label{eqn:totalscalartmunu} \nonumber\\ 
&\qquad - \sum_{k= 0}^\infty\frac{2k+1}{k+1}\alpha_k \left(O^{(1)}O^{(2)}\left(1 +d \beta \left(T^{(1)} + O^{(1)}\right)\right)\left(1 +d \beta \left(T^{(2)} + O^{(2)}\right)\right)\right)^{k+1}\delta^{\mu}_{\phantom{\mu}\nu},
\end{align}
satisfies the combined Ward identity:
\begin{equation}\label{fullWI}
\nabla_\mu T^{\mu}_{\phantom{\mu}\nu} = 0.
\end{equation}
Therefore, $T^{\mu\nu} \equiv T^{\mu}_{\phantom{\mu}\rho} g^{\rho\nu}$ is the energy-momentum tensor of the combined system.

For each pair of scalar operators in the UV and IR theories, we can then have the more general scalar couplings $\beta$ and $\alpha_k$. It should already be evident at this stage how the existence of a local energy-momentum tensor of the full system restricts the hard-soft couplings via the functional forms of the effective sources and background metrics. In practice, for a wide range of energy scales, we should only require a finite number of hard-soft couplings for reasons mentioned previously.

\subsection{Tensorial couplings}
To explore how the effective background metric can be tensorially modified as opposed to being modified by an overall scale factor, we will exploit another identity. Let $g_{\mu\nu}$ and $\tilde{g}_{\mu\nu}$ be two different metric tensors such that one can be smoothly deformed to the other and $z^\mu_{\phantom{\mu}\nu} \equiv g^{\mu\rho}\tilde{g}_{\rho\nu}$. Then we can show that the Levi-Civita connections $\Gamma$ constructed from $g$ and $\tilde\Gamma$ constructed from $\tilde g$ are related by the identity:
\begin{equation}
\tilde{\Gamma}^\rho_{\mu\nu} = \Gamma^\rho_{\mu\nu} + \frac{1}{2}\left(\nabla_\mu \,({\rm \ln \, z})^\rho_{\phantom{\rho} \nu} + \nabla_\nu \,({\rm \ln \, z})^\rho_{\phantom{\rho} \mu}-\nabla^\rho \,({\rm \ln \, z})_{\mu \nu}\right).
\end{equation}
In order to prove this, one can substitute $\tilde{g}_{\mu\nu}$ by  $g_{\mu\nu}+ \delta g_{\mu\nu}$ in the above equation, expand both sides of the equation in $\delta g_{\mu\nu}$, and finally confirm that the identity indeed holds to all orders in this expansion.

Using the above identity, one can show that the Ward identity,
\begin{equation}
\tilde{\nabla}_\mu T^\mu_{\phantom{\mu}\nu} = O \nabla_\nu J,
\end{equation}
in the background $\tilde{g}$ can then be rewritten as
\begin{equation}
\nabla_\mu \left(T^\mu_{\phantom{\mu}\nu} \sqrt{{\rm det}\, z}\right) - \dfrac{1}{2} T^\alpha_{\phantom{\alpha}\beta} \sqrt{{\rm det} \, z} \, \nabla_\nu \left({\rm \ln \, z}\right)^\beta_{\phantom{\alpha} \alpha} - \sqrt{{\rm det} \, z} \, O \, \nabla_\nu J ~ = ~ 0
\end{equation}
in the background metric $g$.

The tensorial couplings then turn out to be given by the following form of the effective sources\footnote{As before, $T^{(i)\mu}_{\phantom{(i)\mu}\nu}= T^{(i)\mu\rho}g^{(i)}_{\rho\nu}$.}
\begin{subequations}
\label{allequations}
\begin{eqnarray}\label{tensor-couple1}
z^{(1)\mu}_{\phantom{(1)\mu}\nu} &= &\exp\left[\left(2\gamma_1\left(T^{(2)\mu}_{\phantom{(2)\mu}\nu}- \dfrac{1}{d} T^{(2)} \delta^\mu_{\phantom{\mu}\nu} \right)+2\gamma_2 T^{(2)\mu}_{\phantom{(2)\mu}\nu}\right)\sqrt{{\rm det}\, z^{(2)}}\right] 
\\\label{tensor-couple2}
z^{(2)\mu}_{\phantom{(2)\mu}\nu} &=& \exp\left[\left(2\gamma_1\left(T^{(1)\mu}_{\phantom{(1)\mu}\nu}- \dfrac{1}{d} T^{(1)}\delta^\mu_{\phantom{\mu}\nu} \right)+2\gamma_2 T^{(1)\mu}_{\phantom{(1)\mu}\nu}\right)\sqrt{{\rm det}\, z^{(1)}}\right] \label{tensor-couple2}\\ \label{tensor-couple3}
J^{(1)} &=&  J^{(2)} \, \, = \, \,0.
\end{eqnarray}
\end{subequations}
In order to define $z^{(i)\mu}_{\phantom{(i)\mu}\nu} \equiv g^{\mu\rho}g^{(i)}_{\rho\nu}$ above, we have chosen a fixed background metric $g$ on which the full system lives. Furthermore, $T^{(i)\mu}_{\phantom{(i)\mu}\nu} \equiv T^{(i)\mu\nu}g^{(i)}_{\mu\nu}$. The above tensorial couplings arise from essentially two available tensor structures, namely the traceless part of the energy-momentum tensor and the energy-momentum tensor itself of the complementary theory, giving rise to the two hard-soft coupling constants $\gamma_{1}$ and $\gamma_2$. We call these couplings tensorial because these do not involve any scalar operator.

The determinants ${\rm det}\, z^{(1)}$ and ${\rm det}\, z^{(2)}$ can be obtained by first evaluating the left and right hand sides of eqs. \eqref{tensor-couple1} and  \eqref{tensor-couple2} which yields:
\begin{eqnarray}
{\rm det}z^{(1)} ~ &= ~ \exp \left[2\gamma_2 T^{(2)}\sqrt{{\rm det}z^{(2)}}\right], \nonumber\\
{\rm det}z^{(2)} ~ &= ~ \exp \left[2 \gamma_2 T^{(1)}\sqrt{{\rm det}z^{(1)}}\right].
\end{eqnarray}
Clearly then,  ${\rm det}\, z^{(1)}$ and ${\rm det}\, z^{(2)}$ are solutions of \footnote{It is easy to check that real and positive solutions of these equations exist at least perturbatively in $\gamma_2$.}
\begin{eqnarray}
{\rm det}z^{(1)} ~ &= ~ \exp \left[2\gamma_2 T^{(2)}\exp \left[\gamma_2 T^{(1)}\sqrt{{\rm det}z^{(1)}}\right]\right], \nonumber\\
{\rm det}z^{(2)} ~ &= ~ \exp \left[2\gamma_2 T^{(1)}\exp \left[\gamma_2 T^{(2)}\sqrt{{\rm det}z^{(2)}}\right]\right]
\end{eqnarray}
These solutions must be substituted in \eqref{tensor-couple1} and \eqref{tensor-couple2} to finally obtain the complete expressions of the effective background metrics as functionals of the energy-momentum tensors of the two sectors.

As a consequence of the above tensorial couplings, we now find that the energy-momentum tensor of the full system takes the form
\begin{align}
T^\mu_{\phantom{\mu}\nu} ~ &= ~ T^{(1)\mu}_{\phantom{(1)\mu}\nu}\sqrt{{\rm det}\, z^{(1)}} ~ + ~ T^{(2)\mu}_{\phantom{(1)\mu}\nu}\sqrt{{\rm det}\, z^{(2)}}\nonumber\\ 
&\quad - \gamma_1\sqrt{{\rm det}\, z^{(1)}}\sqrt{{\rm det}\, z^{(2)}} \left(T^{(1)\alpha}_{\phantom{(2)\mu}\beta}- \dfrac{1}{d} T^{(1)} \delta^\alpha_{\phantom{\mu}\beta}\right)\left(T^{(2)\beta}_{\phantom{(2)\mu}\alpha}- \dfrac{1}{d} T^{(2)} \delta^\beta_{\phantom{\mu}\alpha}\right)\delta^\mu_{\phantom{\mu}\nu} \label{eqn:totaltensortmunu} \nonumber \\
&\quad -\gamma_2\sqrt{{\rm det}\, z^{(1)}}\sqrt{{\rm det}\, z^{(2)}}T^{(1)\alpha}_{\phantom{(2)\mu}\beta}T^{(2)\beta}_{\phantom{(2)\mu}\alpha}\delta^\mu_{\phantom{\mu}\nu}\,  .
\end{align}
satisfying the combined Ward identity \eqref{fullWI} in the fixed background $g$.

\subsection{Combining general scalar and tensorial couplings}
Having independently identified the scalar and tensorial hard-soft couplings, we can put them together to obtain a general class of scalar plus tensorial couplings and the resulting combined energy-momentum tensor of the full theory. In order to do so, we begin by define two functions $U$ and $V$ as below (with $z^{(i)\mu}_{\phantom{(i)\mu}\nu} \equiv g^{\mu\rho}g^{(i)}_{\rho\nu}$):
\begin{equation}\label{eqn:UV}
U ~ := ~ d\beta \left(T^{(1)}+ O^{(1)}\right)\sqrt{{\rm det}z^{(1)}} \quad \text{and} \quad V ~ :=~  d \beta \left(T^{(2)}+ O^{(2)}\right)\sqrt{{\rm det}z^{(2)}} \, .
\end{equation}
Analogous to \eqref{eqn:g1g2}, we impose that the scale factors $e^{d\sigma^{(1)}}$ and $e^{d\sigma^{(2)}}$ in the UV and IR theories should assume the forms:
\begin{equation}
e^{d\sigma^{(1)}} ~ = ~ 1 + e^{-d\sigma^{(2)}} \, V \qquad \text{and} \qquad e^{d\sigma^{(2)}} ~ = ~ 1 + e^{-d\sigma^{(1)}} U \, ,
\end{equation}
Equivalently, we can impose:
\begin{subequations}
\label{allequations}
\begin{align}
e^{d\sigma^{(1)}} ~ &= ~ \dfrac{1+V -U}{2} + \sqrt{U + \left( \dfrac{1+V -U}{2}\right)^2},\label{eqn:sigma1gen} \\ 
e^{d\sigma^{(2)}} ~ &= ~ \dfrac{1+U -V}{2} + \sqrt{V+ \left( \dfrac{1+U -V}{2}\right)^2} \, . \label{eqn:sigma2gen}
\end{align}
\end{subequations}
to relate the scale factors with ${\rm det}\, z^{(1)}$. As an aside, it may be worth noting that the two terms under the square-roots in \eqref{eqn:sigma1gen} and \eqref{eqn:sigma2gen} can be checked to be equal. Finally, we demand that the effective background metrics and sources are given by the expressions:\footnote{As before, $T^{(i)\mu}_{\phantom{(i)\mu}\nu}= T^{(i)\mu\rho}g^{(i)}_{\rho\nu}$.}
\begin{align}\label{tensor-scalar-couple}
z^{(1)\mu}_{\phantom{(1)\mu}\nu} ~ &= ~ \exp\left[\left(2\gamma_1\left(T^{(2)\mu}_{\phantom{(2)\mu}\nu}- \dfrac{1}{d} T^{(2)} \delta^\mu_{\phantom{\mu}\nu} \right)+2\gamma_2 T^{(2)\mu}_{\phantom{(2)\mu}\nu}\right)\sqrt{{\rm det}\, z^{(2)}}\right] e^{2\sigma^{(1)}}, \nonumber\\
z^{(2)\mu}_{\phantom{(2)\mu}\nu} ~ &= ~ \exp\left[\left(2\gamma_1\left(T^{(1)\mu}_{\phantom{(1)\mu}\nu}- \dfrac{1}{d} T^{(1)}\delta^\mu_{\phantom{\mu}\nu} \right)+2\gamma_2 T^{(1)\mu}_{\phantom{(1)\mu}\nu}\right)\sqrt{{\rm det}\, z^{(1)}}\right] e^{2\sigma^{(2)}}, \nonumber\\
J^{(1)} ~ &= ~ \sigma^{(1)}  + \sum_{k = 0}^\infty\alpha_k {O^{(2)}}^{k+1}\left({\rm det}z^{(2)}\right)^{\frac{k+1}{2}} {O^{(1)}}^k \left({\rm det}z^{(1)}\right)^{\frac{k}{2}}, \nonumber \\ 
J^{(2)} ~ &= ~ \sigma^{(2)} + \sum_{k = 0}^\infty\alpha_k {O^{(1)}}^{k+1}\left({\rm det}z^{(1)}\right)^{\frac{k+1}{2}} {O^{(2)}}^k \left({\rm det}z^{(2)}\right)^{\frac{k}{2}}\, ,
\end{align}
where $\sigma^{(1)}$ and $\sigma^{(2)}$ are given by \eqref{eqn:UV}, \eqref{eqn:sigma1gen} and \eqref{eqn:sigma2gen}. As in the case discussed in the previous subsection, we need to solve ${\rm det} z^{(i)}$s self-consistently by evaluating the determinants of the left and right hand sides of the first two equations above.\footnote{Once again we can check that sensible solutions exist at least perturbatively in the hard-soft couplings.} It may readily be checked that when the tensorial couplings are turned off by setting $\gamma_1 = \gamma_2 = 0$, we revert back to the case discussed in Section \ref{sec:scalar-scalar-coupling} where we have obtained:
\begin{align}
\sqrt{{\rm det} z^{(1)}} ~ &= ~ 1+ d\beta \left(T^{(2)} +O^{(2)}\right) \,=\, e^{d\sigma^{(1)}} \qquad \text{and} \nonumber \\
\sqrt{{\rm det} z^{(2)}} ~ &= ~ 1+ d\beta \left(T^{(1)} +O^{(1)}\right)\,=\, e^{d\sigma^{(2)}} \, . \nonumber
\end{align}
The total conserved energy-momentum tensor is given by:
\begin{align} \label{eqn:totalscalar+tensortmunu}
T^\mu_{\phantom{\mu}\nu} ~ &= ~ T^{(1)\mu}_{\phantom{(1)\mu}\nu}\sqrt{{\rm det}\, z^{(1)}} ~ + ~ T^{(2)\mu}_{\phantom{(1)\mu}\nu}\sqrt{{\rm det}\, z^{(2)}} \nonumber\\
&\quad- \gamma_1\sqrt{{\rm det}\, z^{(1)}}\sqrt{{\rm det}\, z^{(2)}} \left(T^{(1)\alpha}_{\phantom{(2)\mu}\beta} - \dfrac{1}{d} T^{(1)} \delta^\alpha_{\phantom{\mu}\beta}\right)\left(T^{(2)\beta}_{\phantom{(2)\mu}\alpha} - \dfrac{1}{d} T^{(2)} \delta^\beta_{\phantom{\mu}\alpha}\right)\delta^\mu_{\phantom{\mu}\nu} \nonumber\\
&\quad - \gamma_2\sqrt{{\rm det}\, z^{(1)}}\sqrt{{\rm det}\, z^{(2)}}T^{(1)\alpha}_{\phantom{(2)\mu}\beta}T^{(2)\beta}_{\phantom{(2)\mu}\alpha}\delta^\mu_{\phantom{\mu}\nu} \nonumber \\
&\quad - \beta \left( T^{(1)} +O^{(1)}\right)\left(T^{(2)} +O^{(2)}\right)\sqrt{{\rm det}\, z^{(1)}}\sqrt{{\rm det}\, z^{(2)}}e^{-d(\sigma^{(1)}+\sigma^{(2)})}\delta^\mu_{\phantom{\mu}\nu} \nonumber\\ &\quad
- \sum_{k= 0}^\infty\frac{2k+1}{k+1}\alpha_k \left(O^{(1)}O^{(2)}\sqrt{({\rm det}z^{(1)})({\rm det}z^{(2)})}\right)^{k+1}\delta^{\mu}_{\phantom{\mu}\nu}.
\end{align}
which satisfies the Ward identity \eqref{fullWI} in the fixed background $g_{\mu\nu}$. We note that for each pair of scalar operators in the UV and IR theories, we can then have the more general tensor and scalar couplings $\gamma_2$, $\gamma_1$, $\beta$ and $\alpha_k$. In the following section, we will present a toy example to demonstrate how we can determine the hard-soft couplings and the parameters of the IR theory as functionals of the perturbative coupling constants. In particular, the sources for irrelevant scalar operators in the IR theory should vanish -- this will play a major role in determining the hard-soft coupling constants. 

Finally, we would like to emphasise that the special phenomenological semi-holographic constructions \cite{Iancu:2014ava,Mukhopadhyay:2015smb,Mukhopadhyay:2016fkl} discussed in Section \ref{review} are special instances of the general hard-soft coupling scheme discussed in this section. These special instances naturally follow if the hard-soft couplings are small and we retain only such leading coupling terms.\footnote{In the model for heavy-ion collisions discussed in Section \ref{review}, we have two pairs of scalar operators. The first pair are the perturbative and \textit{shadow} glueball condensates, and the second pair are the perturbative and \textit{shadow} Pontryagin charge densities.} As discussed before, such phenomenological constructions can be well justified in a certain range of energy scales.

At the end of Section \ref{es}, we will show how the general coupling rules are modified when the full theory couples to external scalar sources. In fact, this investigation will allow us to define all scalar operators in the full theory as appropriate weighted sums of the effective UV and IR operators.

\section{A bi-holographic illustration}\label{sec:holRGapplication}

In this section, we construct a complete toy theory to illustrate the principles of the semi-holographic framework. In our toy theory, we will see how the UV dynamics determines both the hard-soft couplings and the IR theory. We will also see why the bulk fields of the dual IR holographic theory should undergo state-dependent field-redefinitions in the semi-holographic construction although the classical gravitational theory dual to the IR is itself not state-dependent. Furthermore, we will find that $\Lambda\rightarrow\infty$ behaviour of the hard-soft couplings can be obtained from the vacuum but their runnings with the scale can be state-dependent.

The basic simplification in our toy theory consists of replacing the perturbative UV dynamics by a strongly coupled holographic theory admitting a dual classical gravity description on its own. The IR dynamics will be given by a \textit{different} even more strongly coupled holographic theory with a \textit{different} dual classical gravity description. The advantage of this \textit{biholographic} set-up will be that the UV-IR operator map can be simplified by construction. As we will see in the following section, this map will be immensely complex in QCD which is asymptotically free (although we can still proceed systematically). Our bi-holographic construction is designed to establish the conceptual foundations of semi-holography.

\begin{figure}[htbp]
\centering
\includegraphics[clip, trim=2.0cm 9cm 2.0cm 9cm,width=0.85\textwidth]{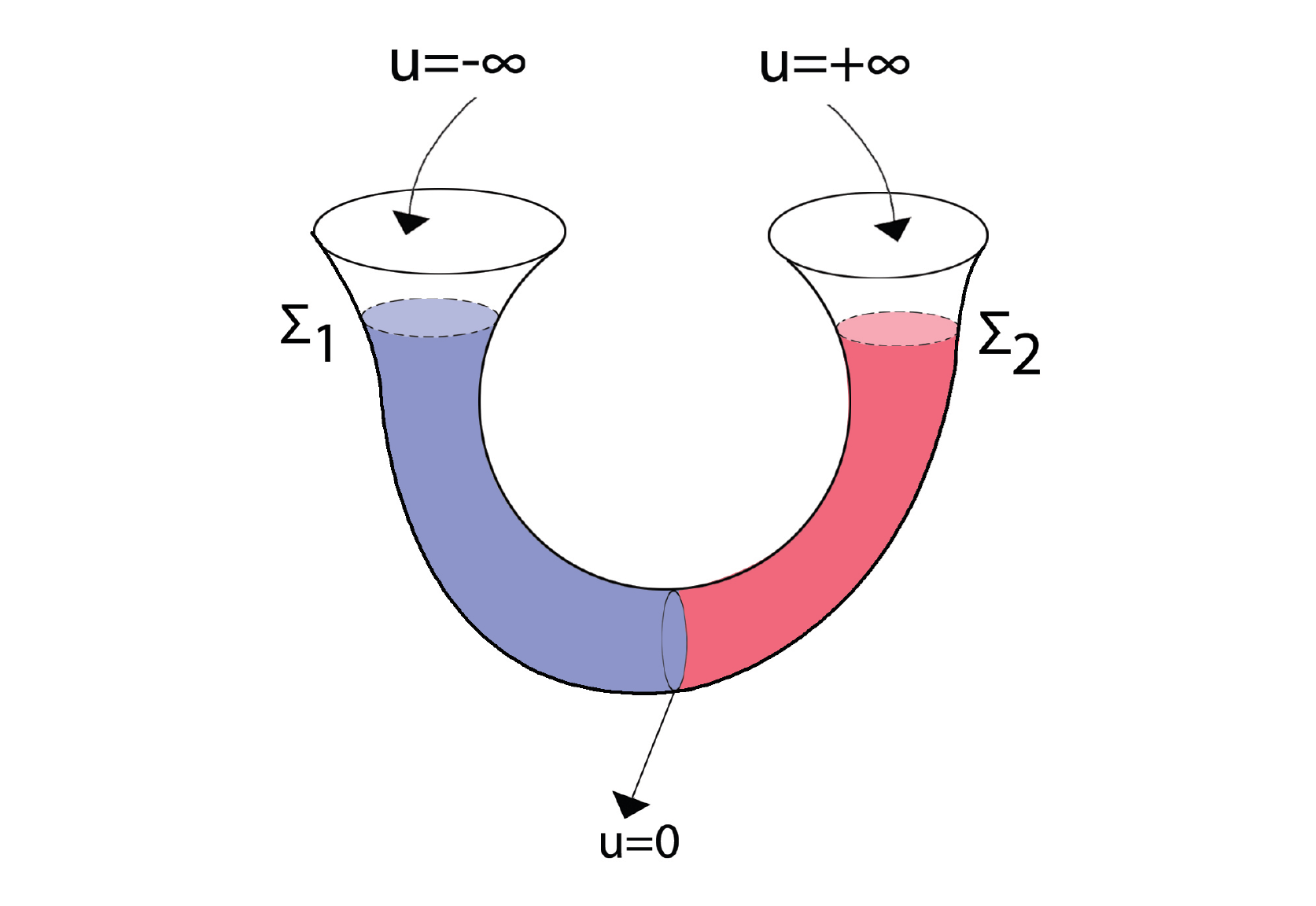}
\caption{Our biholographic toy theory is described by two different holographic UV (blue) and IR (red)  universes with different classical gravity laws. However these are smoothly glued at $u=0$. A scale $\Lambda$ in the full theory in a certain RG scheme is represented by data on the two appropriate hypersurfaces $\Sigma_1$ and $\Sigma_2$ belonging to the two universes as described in the text. In the UV, most contributions come from the UV blue universe. This explains the nomenclature.}
\label{fig:UVandIRUniverses}
\end{figure}

The spirit of our bi-holographic construction is captured in Fig. \ref{fig:UVandIRUniverses}. The UV dynamics is represented by the blue $(d+1)-$dimensional holographic emergent universe which covers the radial domain $-\infty < u <0$ and the IR dynamics is represented by the red $(d+1)-$dimensional holographic emergent universe which covers the radial domain $0 <u < \infty$, with $u$ being the holographic radial coordinate denoting the scale. Each of these geometries is asymptotically $AdS$ and their individual conformal boundaries are at $u= \pm \infty$ respectively. Although the bulk fields are governed by different classical gravity theories in the two different universes, these transit smoothly at the gluing surface $u= 0$. In each universe, we can use the standard rules of holographic duality \cite{Maldacena:1997re,Gubser:1998bc,Witten:1998qj} to extract the effective UV and IR sources and expectation values of the operators from the behaviour of the bulk fields in the respective asymptotic regions. However, the boundary conditions of the two asymptotic regions $u \rightarrow \pm \infty$ are correlated by the general consistent coupling rules of the previous section which leads to the existence of a conserved local energy-momentum tensor of the full dual quantum many-body system.\footnote{This feature distinguishes our construction from those described in \cite{Aharony:2006hz,Kiritsis:2006hy,Kiritsis:2008at} where two or more holographic CFTs are coupled by gluing the boundaries of their dual asymptotically AdS geometries. In our case, the AdS spaces are glued in the interior reflecting that the dual theories \textit{glue} together to form a complete and consistent theory. The AdS boundaries in our case are then coupled non-locally in the sense that the sources specified at the boundaries should be correlated by the rules found in the previous section. Such type of non-local couplings (related to multi-trace couplings of operators in future and past directed parts of the Schwinger-Keldysh contour in the dual theory) have been also recently discussed in \cite{Gao:2016bin} in the context of eternal AdS black holes which have two distinct conformal boundaries. It has been shown that such couplings lead to formation of traversable wormholes leading to a concrete realisation of the ER = EPR conjecture \cite{Maldacena:2013xja} stating that quantum entanglement of degrees of freedom (i.e. Einstein-Podolsky-Rosen pairs) leads to formation of Einstein-Rosen bridges (i.e. wormholes) between distinct space-time regions. In our case, this wormhole is perhaps engineered by our coupling rules as suggested by the construction in \cite{Gao:2016bin} reflecting the entanglement between UV and IR degrees of freedom of the dual system. It is worthwhile to note in this context that although the coupling of the boundaries is non-local,  it is strongly constrained in our construction by the existence of a local and conserved energy-momentum tensor in the full dual many-body system.} 

Crucially, the UV and IR theories cure each other. Individually, the UV and IR universes are singular (in a specific sense to be discussed later) if extended in the regions $u>0$ and $u<0$ respectively. However the smooth gluing at $u=0$ implies that the full holographic construction giving the dynamics of the quantum many-body system has no singularity. Therefore, the IR theory indeed completes the UV theory almost in the same manner in which non-perturbative dynamics cures the Borel singularities of perturbation theory. We can also view the IR Universe extending along $0<u<\infty$ as a second cover the UV universe $-\infty < u < 0$ which thus becomes bi-metric. We will discuss this point of view later.

As mentioned above, the leading asymptotic behaviour of bulk fields giving the effective sources and effective background metric of the respective theories are \textit{coupled}, or rather correlated by the general semi-holographic construction rules established in the previous section. Therefore, the full theory admits a local energy momentum tensor conserved in the actual fixed background metric where all the degrees of freedom live.Here, we will take this fixed background metric to be $\eta_{\mu\nu}$.

At this stage, we should clarify in which sense we are using the terms \textit{UV theory} and \textit{IR theory}. After all, the full energy-momentum tensor constructed in the previous section  receives contributions from both UV and IR theories at any scale. In our case, it means that the microscopic energy-momentum tensor of the dual many-body quantum system will be a complicated combination of the energy-momentum tensors and other data obtained from the sub-leading asymptotic modes of both UV and IR universes. Nevertheless, we will see that the contribution to the the energy-momentum tensor at $\Lambda = \infty$ coming solely from the IR universe is zero in the vacuum state. Furthermore, the scale factors of the effective UV and IR metrics (identified with the boundary metrics of the UV and IR universes) will turn out to be dynamically determined such that the effective UV metric will be slightly compressed and effective IR metric will be slightly dilated compared to the fixed background Minkowski space. This explains our UV and IR nomenclatures. Note these feature are also present in the semi-holographic constructions as discussed before -- the vanishing of the scale-dependent hard-soft couplings in the UV ensures that perturbative contributions dominate in the UV as should be the case in asymptotically free theories like  QCD. In our bi-holographic construction, although the contributions of the UV universe will dominate, the hard-soft couplings will be finite in the limit $\Lambda \rightarrow \infty$.

Furthermore, in our bi-holographic construction, a scale $\Lambda$ in the full theory is represented by the data on the union of two appropriate hypersurfaces $\Sigma_1$ and $\Sigma_2$ in the UV and IR universes respectively in a specific type of RG scheme as shown in Fig. \ref{fig:UVandIRUniverses}. We will discuss this issue in more details later. The IR hypersurface $u= 0$ will represent an endpoint of the RG flow. In fact the geometry near $u= 0$ can be described as an infrared AdS space with zero volume, and so we will argue that it is a fixed point. More generally, however $u= 0$ could be a \textit{wall} representing confinement in the dual theory. 

Conceptually, our bi-holographic construction is thus very different from how RG flow is represented in standard holographic constructions such as that described in \cite{Freedman:1999gp}. In these cases, the full emergent spacetime is described by a \textit{single} gravitational theory and it also has a \textit{single} conformal boundary. Furthermore, although the spacetime has another AdS region in the IR (deep interior), this matches with the rear part of a pure AdS geometry, i.e. the part that contains the Poincare horizon and not the asymptotic conformal boundary. Physically these geometries represent deformation of the UV fixed point in the dual field theory by a relevant operator as a result of which it flows to a different IR fixed point -- a geometric $c-$function can also be constructed \cite{Freedman:1999gp} reproducing the central charges of the UV and IR fixed points. Our bi-holographic construction however should not be thought of as a flow from an UV to an IR fixed point driven by a relevant deformation.\footnote{The general semi-holographic construction can of course apply to such a case particularly if the UV fixed point is weakly coupled.} Rather the UV region in our case represents a strongly coupled version of usual perturbative dynamics, and the IR region represents the non-perturbative sector that exists as a \textit{shadow} of the perturbative degrees of freedom in many-body quantum systems. In our case, the shadow IR theory and the hard-soft coupling constants will be determined by the parameters of the UV gravitational theory via:
\begin{enumerate}
\item the general coupling rules of the previous section ensuring the existence of a conserved energy-momentum tensor of the full system, 
\item the vanishing of the sources for irrelevant IR operator(s), and
\item the continuity of bulk fields and their radial derivatives up to appropriate orders at the matching hypersurface $u=0$.
\end{enumerate}

\subsection{A useful reconstruction theorem}\label{reconstruction}
We can readily proceed with some simplifying assumptions. The first assumption is that both the UV and IR holographic classical gravity descriptions are provided by Einstein-dilaton theories consisting of a scalar field with (different) potentials and minimally coupled to gravity. The gravitational theories are then individually described by the respective actions:
\begin{equation}\label{eqn:action}
S^{\rm UV, IR}_{\rm grav} ~ = ~ \dfrac{1}{16\pi G_N}\int {\rm d}^{d+1}x\,\sqrt{-G} \left(R - G^{MN} \partial_M\Phi \partial_N \Phi - 2 V^{\rm UV, IR}(\Phi) \right),
\end{equation}
with $d$ denoting the number of spacetime dimensions of the dual quantum many-body system. Note $\Phi$ is dimensionless in the above equation. 

To describe the dual vacuum state, it will be convenient to use the \textit{domain-wall coordinates} in which the bulk metric and scalar fields assume the forms:\footnote{It has been observed that the domain-wall radial coordinate $u$ can be directly related to the energy-scale of the dual theory \cite{Erdmenger:2001ja,Kiritsis:2014kua,Behr:2015aat}. }
\begin{equation}\label{eqn:metricansatzdomainwall}
{\rm d}s^2 ~ = ~ {\rm d}u^2 + e^{2\rho(u)} \eta_{\mu\nu}{\rm d}x^\mu {\rm d}x^\nu, \qquad \Phi = \Phi(u).
\end{equation}
The conformal boundaries are at $u=\pm\infty$. We choose $d=4$. In the domain-wall coordinates, the gravitational equations of motion for $\rho$ and $\Phi$ in the domain $-\infty < u < 0$ can be written in the form:
\begin{eqnarray}
\rho'' &=& -\frac{1}{3}{\Phi'}^2,\label{graveqn1} \\ 
\frac{3}{2}\rho'' + 6{\rho'}^2 &=& -V^{\rm UV}(\Phi) \label{graveqn2} .
\end{eqnarray}
The equations of motion for $\rho$ and $\Phi$ in the domain $ 0<u<\infty$ are exactly as above but with $V^{\rm UV}$ replaced by $V^{\rm IR}$.

Instead of choosing $V^{\rm UV}(\Phi)$ and $V^{\rm IR}(\Phi)$ in order to specify the UV and IR gravitational theories, we will take advantage of the so-called \textit{reconstruction theorem} which states that there exists a unique map between a choice of the radial profile of the scale factor, i.e. $\rho(u)$ and the bulk scalar potential $V(\Phi)$ which supports it.\footnote{As far as we are aware of, this theorem was first stated in the context of cosmology in \cite{Tsamis:1997rk}.} The proof of this theorem is straightforward. Suppose we know $\rho(u)$. We can then first use eq. \eqref{graveqn1} to construct $\Phi(u)$. However, this requires an integration constant. If the spacetime has a conformal boundary, at $u= -\infty$ for instance, the behaviour of $\rho$ near $u= -\infty$ should be as follows:
\begin{eqnarray}
\rho(u) = \rho_0 - \frac{u}{L} - \rho_{\delta} \exp\left(2\delta\frac{u}{L}\right) + {\rm subleading \,\, terms},
\end{eqnarray}
with $\delta  > 0$ and $\rho_\delta > 0$. The latter restriction results from the requirement that $\rho''$ should be negative in order for a solution for $\Phi$ to exist as should be clear from \eqref{graveqn1}. One can readily check from \eqref{graveqn1} that the asymptotic behaviour of $\Phi$ should be:
\begin{equation}\label{Phir}
\Phi(u) = \Phi_0 \pm 2\sqrt{3 \rho_\delta}  \exp\left(\delta\frac{u}{L}\right) + {\rm subleading \,\, terms}.
\end{equation}
If $\delta \neq d$, then $2\sqrt{3 \rho_\delta}$ should correspond to the non-normalisable mode or the normalisable mode of $\Phi$. The constant $\Phi_0$ is merely an integration constant. 

We can then invert this function $\Phi(u)$ to obtain $u(\Phi)$. Furthermore, from eq. \eqref{graveqn2}, we can readily obtain $V(u)$ substituting $u(\Phi)$ in which yields $V(\Phi)$. Thus we construct the $V(\Phi)$ corresponding to a specific $\rho(u)$. This ends the proof of the reconstruction theorem. Note this proof assumes that the inverse function $u(\Phi)$ exists. We have to ensure that this is indeed the case.  

The crucial point is that one can readily see from \eqref{graveqn2} that asymptotically (i.e. near $u= \infty$ or equivalently near $\Phi \approx \Phi_0$), $V(\Phi)$ should have the expansion
\begin{equation}
V(\Phi) = - \frac{6}{L^2} + \frac{1}{2}m^2 (\Phi - \Phi_0)^2 + \mathcal{O}(\Phi - \Phi_0)^3,
\end{equation}
when $d=4$ with
\begin{equation}
m^2 L^2 = \delta (\delta -4).
\end{equation}
This implies that $V(\Phi)$ should have a critical point at $\Phi = \Phi_0$ in order that the geometry can become asymptotically anti-de Sitter. Furthermore, the field $\Phi - \Phi_0$ and not $\Phi$ corresponds to the dual operator $O$ with scaling dimension $\delta$ or $4- \delta$ when $\delta \neq 4$. Therefore, without loss of generality when $\delta \neq 4$, we can always employ the field redefinition $\Phi \rightarrow \Phi - \Phi_0$ and set the integration constant to be zero. When $\delta = 4$, $\Phi$ is massless and we need to use the holographic correspondence to figure out what the integration constant should be since $\Phi_0$ then corresponds to a marginal coupling of the dual field theory. Usually, it is put to zero even in this case. We will not deal with the massless scalar case here.

\subsection{The bi-holographic vacuum}
We first focus on constructing the bi-holographic vacuum state. Let us begin by individually choosing an ansatz for the UV and IR gravitational theories. We take the advantage of the reconstruction theorem described above by making an ansatz for $\rho(u)$ in the respective domains instead of doing so for the potentials $V^{\rm UV,IR}(\Phi)$. For the sake of convenience we also put $8 \pi G_N = 1$. We will set $d=4$. 

Since both the UV and IR gravitational theories have the same $G_N$ as clear from \eqref{eqn:action} and their AdS radii will turn out to be of the same order if not equal, we can take the large $N$ limit in both sectors simultaneously. Therefore, we can not only suppress quantum gravity loops in each gravitational theory, but also hybrid ones. This justifies our assumption that quantum gravity effects can be ignored in both gravitational theories. 

\subsubsection{The UV domain} 
To take advantage of the reconstruction theorem, we choose the scale factor profile in the (dual UV) domain $-\infty < u <0$ to be:
\begin{equation}\label{rhoUV}
\rho^{\text{UV}}\left(u\right) ~ = ~ A_0 - \dfrac{u}{L^{\text{UV}}} - A_1 \tanh\left( \dfrac{u}{L^{\text{UV}}}\right) + A_2 \tanh\left(2 \dfrac{u}{L^{\text{UV}}}\right). 
\end{equation}
We can motivate the choice of the $\tanh$ functions as follows. These lead to right exponentially subleading asymptotic behaviour at $u=-\infty$ and leads to $\rho'' = 0$ at $u= 0$. Crucially, if we choose our parameters $A_1$ and $A_2$ such that $\rho'' < 0$ for $u< 0$, then typically $\rho'' > 0$ should follow for $u > 0$. This does not lead to a curvature singularity, however \eqref{graveqn1} implies that $\Phi$ has no real solution for $u > 0$, i.e. it signals the end of UV spacetime at $u=0$. This leads to a singularity in the sense of \textit{geodetic incompleteness}, because any freely falling observer can reach the edge $u=0$ from a finite value of $u$ in finite proper time. This singularity is eventually cured by the emergence of the IR universe.

The above choice for $\rho^{\rm UV} (u)$ leads to a unique solution for $\Phi^{\rm UV}(u)$ whose asymptotic (i.e. $u\rightarrow -\infty$) behaviour is:
\begin{equation}
\Phi^{\text{UV}}\left(u\right) ~ = ~ U_1 \exp\left(\dfrac{u}{L^{\text{UV}}}\right) + \cdots + V_1 \exp\left(3 \dfrac{u}{L^{\text{UV}}}\right) + \cdots \label{eqn:reluvphi} \, ,
\end{equation}
with\footnote{$U_1$ is determined up to a sign. We make a choice of sign here.} 
\begin{equation}
U_1 = -2 \sqrt{6} \sqrt{A_1} \,, \quad V_1 = 4 \sqrt{\dfrac{2}{3}} \dfrac{A_1 + A_2}{\sqrt{A_1}}.
\end{equation}
We can readily obtain $V^{\rm UV}(\Phi)$ too from \eqref{graveqn2} as described above but it's complete explicit form will not be of much importance for us. The only information in $V^{\rm UV}(\Phi)$ which is significant for us is that the mass of $\Phi^{\text UV}$ which is given by $m^2{L^{\rm UV}}^2 = -3 $ which also implies that the corresponding operator $O^{(1)}$ has scaling dimension $\Delta^{\rm UV} = 3$ at the UV fixed point, and is therefore a relevant operator. Furthermore, $U_1 \neq 0$ implies that the UV fixed point is subjected to a relevant deformation and $U_1 {L^{\rm UV}}^{-1}$ is the relevant coupling\footnote{Although the full theory is not a relevant deformation of a UV fixed point as discussed before, the UV holographic theory individually can be described in such terms. This is owing to the fact that the holographic geometry will be asymptotically AdS.} as will be clear once we transform to the Fefferman-Graham coordinates. The Fefferman-Graham radial coordinate $z$ in which the metric assumes the form:
\begin{equation}
{\rm d}s^2 = \left(\frac{{L^{\rm UV}}^2}{z^2}\right)\left({\rm d}z^2+e^{\tilde{\rho}^{\rm UV}(z)}\eta_{\mu\nu}{\rm d}x^\mu {\rm d}x^\nu\right)
\end{equation}
is related to $u$ by $z = L^{\rm UV} \exp (u/L^{\rm UV})$. In the Fefferman-Graham coordinates we obtain:
\begin{eqnarray}
 e^{2 \tilde{\rho}^{\text{UV}}(z)} &=& e^{2 \left(A_0 + A_1 - A_2\right)}  \left(1 - 4 A_1 \frac{z^2}{{L^{\rm UV}}^2} + {4} \left(A_1 + 2 A_1^2 + A_2 \right)\frac{z^4}{{L^{\rm UV}}^4} + \cdots \right),\nonumber\\
 \Phi^{\text{UV}}\left(z\right) &=&   -2 \sqrt{6} \sqrt{A_1}  \frac{z}{L^{\rm UV}} + 4 \sqrt{\dfrac{2}{3}} \dfrac{A_1 + A_2}{\sqrt{A_1}} \frac{z^3}{{L^{\rm UV}}^3}+ \cdots \, .
\end{eqnarray}
We can readily perform holographic renormalisation in the Fefferman-Graham coordinates to extract the sources and the expectation values of the operators in the UV description \cite{Henningson:1998ey,Balasubramanian:1999re,deBoer:1999tgo,deHaro:2000vlm,Skenderis:2002wp}. The scale factor $\sigma^{(1)}$ in the boundary metric $g^{(1)}_{\mu\nu}= e^{2\sigma^{(1)}}\eta_{\mu\nu}$ and the source $J^{(1)}$ for the scalar operator $O^{(1)}$ are given by:
\begin{eqnarray}\label{sourceUV}
\sigma^{(1)} = A_0 + A_1 - A_2, \quad J^{(1)} = -2 \sqrt{6} \sqrt{A_1}{L^{\rm UV}}^{-1}.
\end{eqnarray}
The expectation values of the trace of the energy-momentum tensor $T^{(1)}$ (defined as $T^{(1)} := T^{(1)\mu\nu}g^{(1)}_{\mu\nu}$ as in the previous section) and $O^{(1)}$ are given by:\footnote{We have used the minimal subtraction scheme in which we do not obtain any new parameter from the regularisation procedure as no finite counterterm is invoked. In this case, the scheme dependence arises only from a finite counterterm proportional to ${J^{(1)}}^4$. It is dropped in the minimal subtraction scheme. There is a beautiful independent justification of the minimal subtraction scheme \cite{Anselmi:2000fu,Erdmenger:2001ja} (see also \cite{Behr:2015aat} for an yet another perspective) as only in this scheme one can define holographic $c-$function and beta functions which satisfy identities analogous to those in the field-theoretic local Wilsonian RG flows constructed by Osborn \cite{Osborn:1991gm}.}
\begin{eqnarray}\label{vevUV}
T^{(1)} = 16 \left(A_1 +A_2\right) {L^{\rm UV}}^{-4}, \quad O^{(1)} = - 4 \sqrt{\dfrac{2}{3}} \dfrac{A_1 + A_2}{\sqrt{A_1}}{L^{\rm UV}}^{-3}.
\end{eqnarray}
As a consistency check, we note that the CFT Ward identity $T^{(1)} = J^{(1)} O^{(1)}$ which is scheme-independent is indeed satisfied by the above values. We also note that when $\Delta = 3$, we also have the possibility of alternative quantisation in which case the field $\Phi$ corresponds to an operator with $\Delta = 1$ (which saturates the unitarity bound on lowest possible dimensions of scalar primary operators in a CFT) and the roles of $J^{(1)}$ and $O^{(1)}$ can be interchanged. Here, we perform the more usual quantisation.

\subsubsection{The IR domain} 
In the infrared domain $0 < u < \infty$, we choose the scale factor $\rho^{\rm IR}(u)$ to be:
\begin{equation}\label{rhoIR} 
\rho^{\text{IR}}\left(u\right) =  B_0 - \dfrac{u}{L^{\text{IR}}} - B_1 \tanh\left(5 \dfrac{u}{L^{\text{IR}}}\right) + B_2 \tanh\left(10  \dfrac{u}{L^{\text{IR}}}\right),
\end{equation}
with $L^{\rm IR} < 0$ so that the conformal boundary is indeed at $u = \infty$. The choice of $tanh$ functions can be motivated by similar arguments presented in the UV case -- we need an edge singularity at $u= 0$ which is cured by the gluing to the UV universe.  This choice then implies that $\Phi(u)$ has the asymptotic expansion:\footnote{Once again $V_2$ is determined up to a sign. We make a choice of sign here.}
\begin{equation}\label{eqn:irrelirphi}
\Phi^{\text{IR}}\left(u\right) =  V_2 \exp\left(5 \dfrac{u}{L^{\text{IR}}}\right) + \cdots \,, \quad \text{with} \quad V_2 =  -2\sqrt{6}\sqrt{B_1}.
\end{equation}
Therefore $\Phi$ in the IR region corresponds to an irrelevant operator with $\Delta = 5$. We can now define a new Fefferman-Graham coordinate via $u = L_{\text{IR}} \log\left(\tilde{z}\right)$ suitable for the IR asymptotia which is at $\tilde{z}=0$. The asymptotic expansions are:
\begin{align}
\tilde{z}^2 e^{2 \rho_{\text{IR}}\left(\tilde{z}\right)} ~ &= ~ e^{2 \left(B_0 + B_1 -B_2 \right)} \left(1 - 4 B_1\frac{\tilde{z}^{10}}{{L^{\rm IR}}^{10}}  + \cdots\right) \\
\phi_{\text{IR}}\left(\tilde{z}\right) ~ &= ~ V_2 \frac{\tilde{z}^5}{{L^{\rm IR}}^5} + \cdots \, .
\end{align}
With our choice of $\rho$ in the IR theory, the scalar source is vanishing while the scalar vev is parametrised by $V_2$. Indeed if the scalar source would not have vanished, it would have lead to a runaway asymptotic behaviour causing a curvature singularity. As the dual operator is irrelevant, its source should vanish as otherwise we cannot find the corresponding state in the holographic correspondence as well. The effective IR metric, which is the boundary metric of the IR asymptotic region is given by 
$e^{2 \left(B_0 + B_1 - B_2 \right)} \eta_{\mu\nu}$, while the IR stress-tensor is vanishing. Thus, we obtain
\begin{equation}\label{sourceIR}
\sigma^{(2)} = B_0 + B_1 - B_2, \quad J^{(2)} = 0,
\end{equation}
and 
\begin{equation}\label{vevIR}
T^{(2)} = 0, \quad O^{(2)} = 2\sqrt{6}\sqrt{B_1}{L^{\rm IR}}^{-5}.
\end{equation}

\subsubsection{Gluing and determining the full theory}
For the full construction, we need to consider the hard-soft couplings. We make a simplistic assumption that the tensorial hard-soft couplings $\gamma_1$ and $\gamma_2$ are zero. We also make another assumption that we can set all scalar hard-soft couplings $\alpha_k$ to zero except for $\alpha_0$. Therefore, $\alpha_0$ and $\beta$ are the only non-vanishing hard-soft couplings in our construction. In what follows, we will denote $\alpha_0$ by $\alpha$ for notational convenience. Our coupling rules thus (resulting from setting $\alpha_k = 0$ for $k \neq 0$ and $d =4$ in  \eqref{eqn:J1} and \eqref{eqn:J2}) are:
\begin{subequations}
\label{allequations}
\begin{align}
e^{4\sigma^{(1)}} ~ &= ~ 1 +4 \beta \left(T^{(2)} + O^{(2)}\right) ,\label{eqn:sigma1n}\\
e^{4\sigma^{(2)}} ~ &= ~ 1+ 4 \beta  \left(T^{(1)} + O^{(1)}\right), \label{eqn:sigma2n}\\
J^{(1)} ~ &= ~ \frac{1}{4}{\rm ln} \left(1 +4\beta\left(T^{(2)} + O^{(2)}\right)\right) + \alpha O^{(2)},\label{eqn:J1n}\\
0~ &= ~ \frac{1}{4}{\rm ln} \left(1 + 4\beta\left(T^{(1)} + O^{(1)}\right)\right) + \alpha O^{(1)}\, . \label{eqn:J2n}
\end{align}
\end{subequations}
In the final equation above we have used $J^{(2)} = 0$, i.e. the source of the irrelevant IR operator must vanish.
 
Furthermore, we impose that the UV and IR geometries can be smoothly glued along their edges which coincide at $u=0$. This matching should cure the respective \textit{edge} singularity (resulting from geodetic incompleteness) as discussed above. Therefore, the metric and the bulk scalar field, and also their radial derivatives up to appropriate orders should be continuous at $u= 0$. There is one subtle point we need to take into account during the gluing procedure. The asymptotic region $u = \infty$ in the IR domain naturally corresponds to UV rather than IR, however in the full theory it represents IR contributions. It is then natural to reverse the scale (radial) orientation of the IR geometry while gluing it to the UV geometry at $u=0$. Equivalently, we should set $\rho$ to $-\rho$ in the IR geometry before we glue it to the IR. One can then also think that the $\rho$ travels back to $-\infty$ from $0$, so that the IR geometry gives another cover of the spacetime whose full extension is $-\infty < u <0$ and an observer simply can pass smoothly from the UV cover to the IR cover. The spacetime is thus bi-metric. The smooth gluing of the two Universes is ensured if at $u = 0$:\footnote{A more diffeomorphism invariant statement is that on the hypersurface $u=0$, the induced metric obtained from the UV and IR sides should match, and the Brown-York stress tensors should also match but with a flipped sign of the IR term. Such a type of gluing with reversed orientation of one manifold has also been considered in \cite{Skenderis:2008dg} in the context of constructing holographic bulk analogue of Schwinger-Keldysh time contour (which reverses and flows back in time.)}
\begin{equation}
\rho^{\rm UV} = \rho^{\rm IR}, \quad {\rho^{\rm UV}}' = -{\rho^{\rm IR}}', \quad \Phi^{\rm UV} = \Phi^{\rm IR}, \quad {\Phi^{\rm UV}}' = -{\Phi^{\rm IR}}'.
\end{equation}
By our choices of $\rho$, $\rho'' = 0$ at $u= 0$ whether we approach from the UV side or the IR side and therefore we automatically obtain from \eqref{graveqn1} that $\Phi' = 0$ from both ends and is hence continuous. Effectively we thus have only two matching conditions, namely
\begin{equation}\label{matching}
\rho^{\rm UV} = \rho^{\rm IR}\quad \text{and} \quad{\rho^{\rm UV}}' = -{\rho^{\rm IR}}' \quad \text{at $u=0$.}
\end{equation}
The matching of $\Phi$ is ensured via a field redefinition. As discussed in Section \ref{reconstruction}, we can always redefine $\Phi$ as $\Phi - \Phi_0$ with $\Phi_0$ being the asymptotic value of $\Phi$  (which after the redefinition becomes zero) so that the potential $V(\Phi)$ has no tadpole term at $\Phi = 0$. However, if there are two asymptotic boundaries, as in our construction, we can do this redefinition in one asymptotic region only. In this case, the integration constant $\Phi_0$ in the other asymptotic region should be set by continuity. We will choose $\Phi_0 = 0$ in the IR end and obtain the value of $\Phi_0$ at the UV end. We need to check that $\Phi_0$ which can be obtained by integrating \eqref{graveqn1} should be finite. This will indeed be the case.

It is fairly obvious that $L^{\rm UV}$ sets the dimensions of all dimensionful parameters in the field theory including $\alpha$ and $\beta$, the hard-soft couplings. Without loss of generality, we can set $L^{\rm UV} = 1$. The (dimensionless) parameters which determine our UV theory are $A_0$, $A_1$ and $A_2$. However, $A_0$ only contributes to the scale factor ($\sigma^{(1)}$) of the effective metric of the UV theory and does not play any role in determining $V^{\rm UV}$. So we can regard $A_1$ and $A_2$ as the true UV parameters. The other parameters are $\alpha$ and $\beta$ (the hard-soft couplings) which are dimensionful and $\delta \equiv L^{\rm IR}/L^{\rm UV}$, $B_0$, $B_1$ and $B_2$ (determining the IR theory) which are dimensionless. Including all parameters of the UV and IR theories and the hard-soft couplings we have in total 9 parameters.

The set of parameters should be such that we must satisfy the 4 coupling equations \eqref{eqn:sigma1n}, \eqref{eqn:sigma2n},\eqref{eqn:J1n} and \eqref{eqn:J2n}, and the 2 matching equations in \eqref{matching}. Since our 9 parameters should satisfy 6 equations, we can determine 6 of our parameters from 3. We choose the 3 parameters which determine the rest to be the UV parameters $A_0$, $A_1$ and $A_2$. In practice, it is easier to choose $A_1$, $B_2$ and $\delta$ which gives the ratio of the UV and IR scales instead as the set of independent parameters. Nevertheless, we can check that it is equivalent to making the right choices for $A_i$s and then determining $B_2$ and $\delta$. We will proceed by choosing $A_1 = B_2 = 1$ and $\delta = -4.91$ (recall that our parametrisation \eqref{rhoIR} require $L^{\rm IR}$ to be negative). $\vert \delta \vert > 1$ implies that the IR theory is more strongly coupled.

It is convenient to first utilise the matching equations \eqref{matching} to obtain:
\begin{equation}
\label{gluing1}
B_0 ~ = ~ A_0  \, , ~~ B_1 ~ = ~ - \dfrac{1}{5}  \left(1 + 10 B_2 + \delta \left(1+ A_1 - 2 A_2\right)\right).
\end{equation}

We then utilise \eqref{eqn:sigma2n} and \eqref{eqn:J1n} to note that $\alpha$ and $\beta$ should be given by:
\begin{equation}\label{a&b}
\alpha = \frac{4J^{(1)} -\ln(1 + 4\beta(T^{(2)}+O^{(2)}))}{4O^{(2)}(1 + 4\beta(T^{(2)}+O^{(2)}))}, \quad \text{and} \quad \beta = \frac{e^{4\sigma^{(2)}}-1}{4(T^{(1)} + O^{(1)})}.
\end{equation}
The right hand sides above are given by the parameters of the UV and IR theory via \eqref{sourceUV}, \eqref{vevUV}, \eqref{sourceIR} and \eqref{vevIR}. Therefore we obtain $\alpha$ and $\beta$ in terms of other parameters, namely $A_i$s, $B_i$s and $\delta$. Since $B_0$ and $B_1$ are given by \eqref{gluing1}, and the values of $A_1$, $B_2$ and $\delta$ have been fixed, $\alpha$ and $\beta$ are now functions of $A_0$ and $A_2$. 

Substituting \eqref{gluing1} and \eqref{a&b} in \eqref{eqn:sigma1n} and \eqref{eqn:J2n}, and using the fixed values of $A_1$, $B_2$ and $\delta$, we can determine the values of $A_0$ and $A_2$ numerically. These numerical values are then used to obtain $B_0$ and $B_1$ from \eqref{gluing1}. Finally, we can use  \eqref{a&b} to determine $\alpha$ and $\beta$. 

Doing so, we obtain $A_2 = -0.25$. As discussed above, we can now also claim that we have actually set $A_1 = 1$, $A_2 = -0.25$ and $\delta = -4.91$ and have determined all other parameters in terms of these. In the end, we obtain $A_0 = -1.25$, $B_0 =-1.25 $, $B_1 = 0.25$ and $B_2 = -1$. Furthermore, the hard-soft couplings in units $L^{\rm UV} = 1$ are 
\be
\alpha = 5.7\times 10^3, \quad \text{and} \quad \beta = 1.2\times 10^{-4}. 
\ee
Of course we have determined these values of $\alpha$ and $\beta$ in the limit $\Lambda \rightarrow \infty$ of the dual field-theoretic system. It is indeed a bit surprising that $\alpha$ is so enormously large and $\beta$ is so tiny. This completes determining all parameters of the IR theory and hard-soft couplings in terms of the dimensionless UV parameters $A_0$, $A_1$ and $A_2$. 

It is also interesting to note that as a result of our solutions we obtain 
\begin{equation}
\sigma^{(1)} = -1.03 \times 10^{-7}, \quad \text{and} \quad\sigma^{(2)} =1.16 \times 10^{-3}.
\end{equation}
This implies the effective UV metric is slightly compressed and the effective IR metric is slightly dilated compared to the background flat Minkowski space as claimed before. 

Finally the other non-vanishing effective sources and vevs in units $L^{\rm UV} = 1$ turn out to be:
\begin{equation}
T^{(1)} = 12.06, \quad J^{(1)} = -2 \sqrt{6},  \quad O^{(1)} = -2.46, \quad O^{(2)} = -8.6 \times 10^{-4}.
\end{equation}
It is also reassuring to see that the effective IR vev is small in units $L^{\rm UV} = 1$ compared to the effective UV vev.  

\begin{figure}[ht]
\centering
\includegraphics[width=0.8\textwidth]{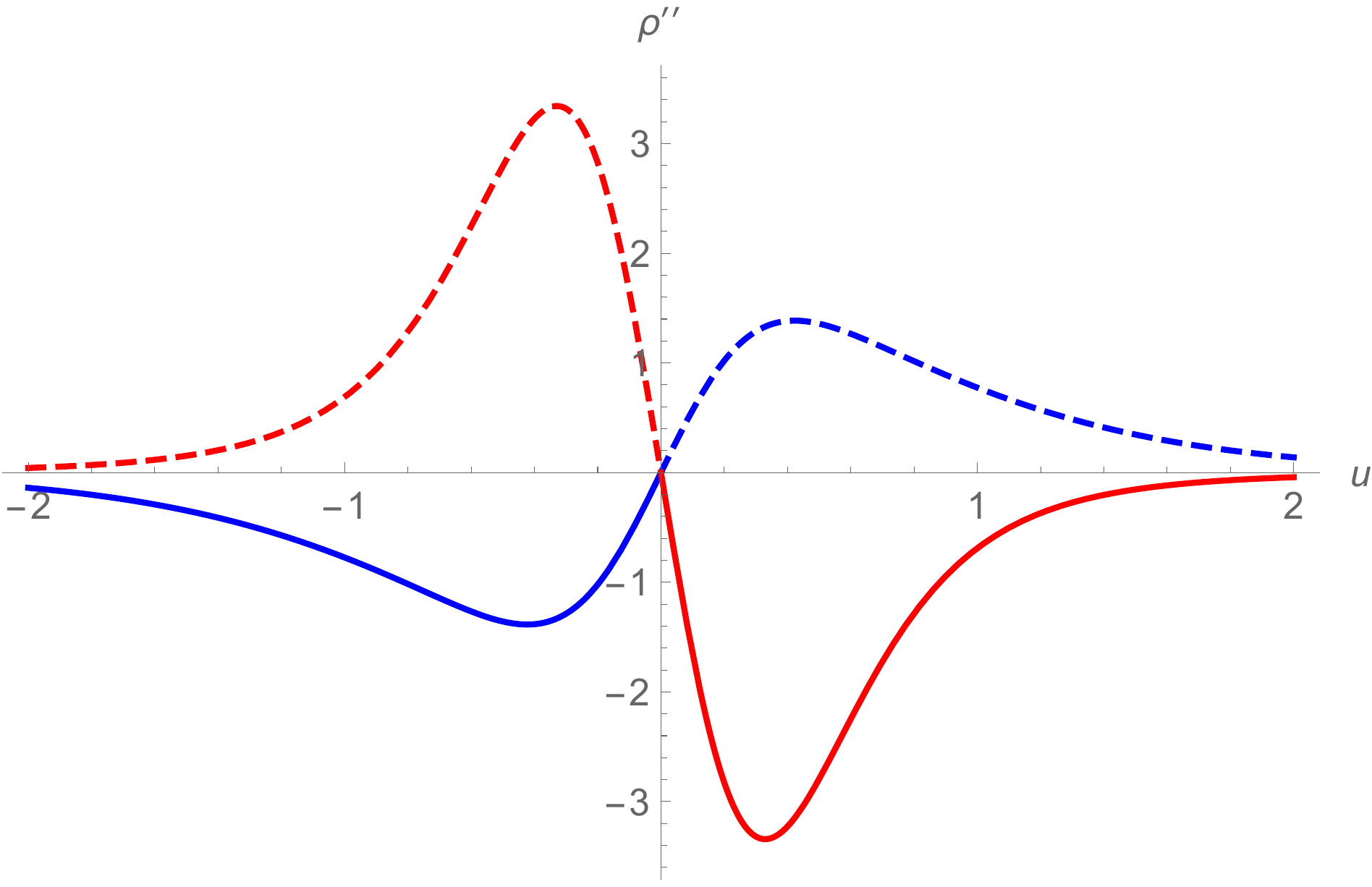}
\caption{\label{fig:rhopp}Plot of $\rho''\left(u\right)$; the blue curve refers to the UV while the red one refers to the IR. The dashed lines indicate the regions where $\rho'' > 0$ -- here the solutions to $\Phi$ do not exist and hence these regions should be discarded. It is clear then how the gluing cures the edge singularities arising from the geodetic incompleteness in each individual component Universe.}
\end{figure}

The most interesting feature is the behaviour of $\rho''$ which has been plotted in Fig. \ref{fig:rhopp}. It is clear from the figure that the gluing cures the edge singularities of each component Universe arising from the geodetic incompleteness -- if extended to $u>0$ and $u<0$, $\rho''$ becomes positive in the UV and IR universes respectively.

\begin{figure}[ht]
\begin{minipage}[b]{0.45\linewidth}
\centering
\includegraphics[width=\textwidth]{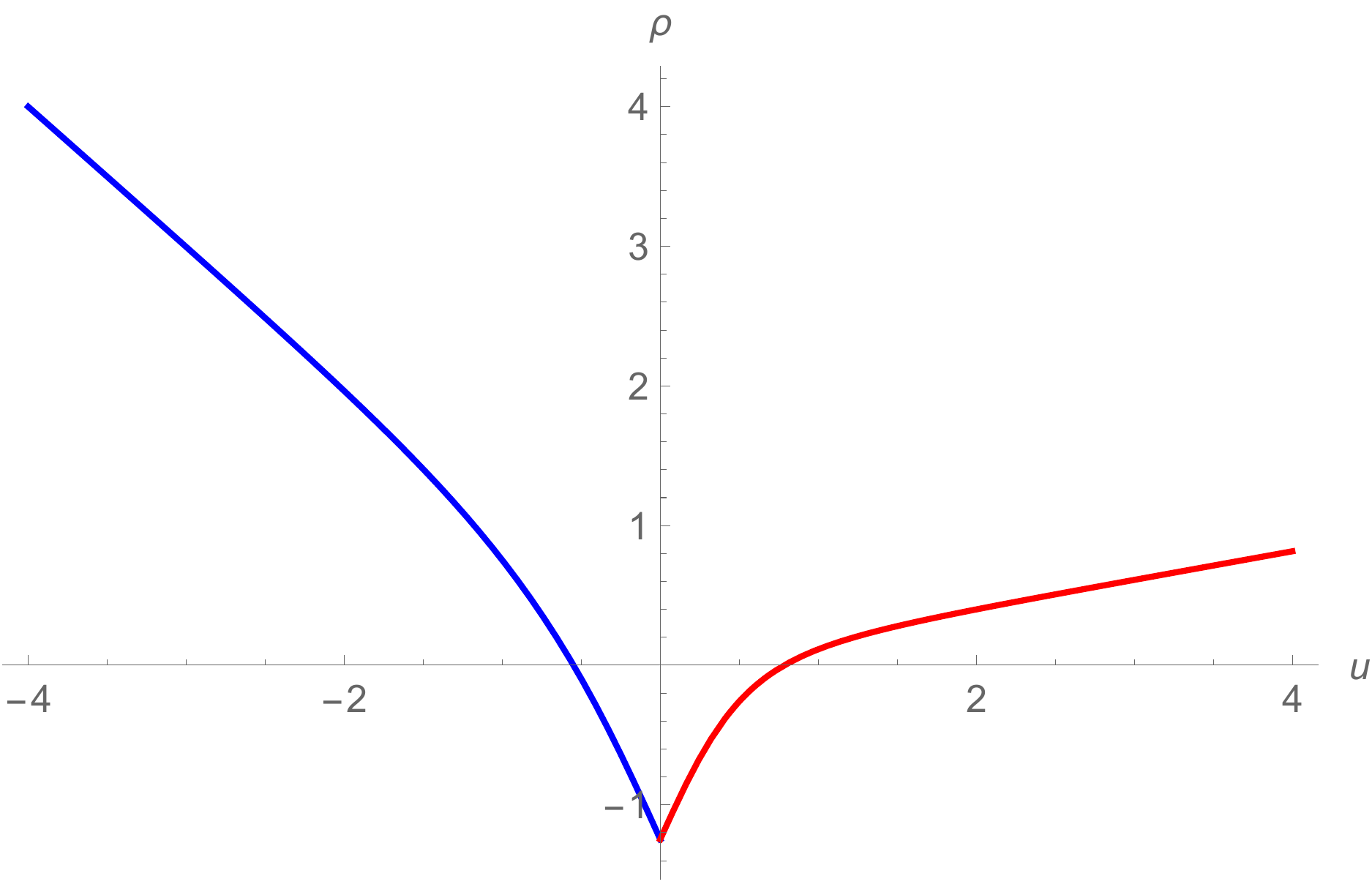}
\caption{\label{fig:rhoofu}Plot of the $\rho\left(u\right)$; the blue curve refers to the UV region while the red one refers to the IR region. $\rho'$ is continuous at $u=0$ if we flip the sign of $\rho'$ on the IR side.}
\end{minipage}
\hspace{0.5cm}
\begin{minipage}[b]{0.45\linewidth}
\centering
\includegraphics[width=\textwidth]{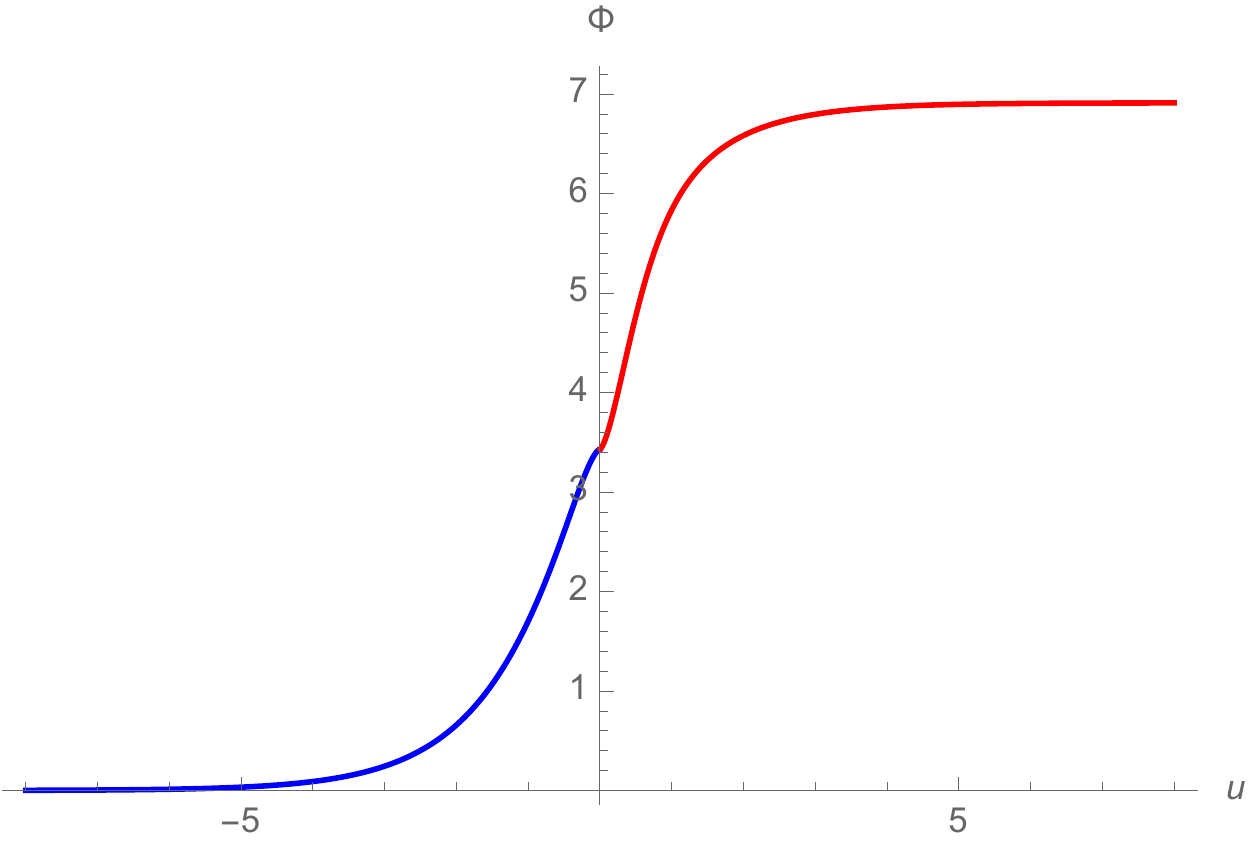}
\caption{\label{fig:phiofu}Plot of the scalar field $\phi\left(u\right)$; the blue curve refers to the UV region while the red one refers to the IR region. Note that $\Phi'=0$ at $u= 0$ where $\Phi \approx 3.43$. The integration constant $\Phi_0$ (see text) at $u=\infty$ is simply $\Phi (u= \infty)$ which is approximately 6.91.}
\end{minipage}
\end{figure}

We plot the scale factor $\rho\left(u\right)$ in Fig. \ref{fig:rhoofu} and $\Phi\left(u\right)$ in Fig. \ref{fig:phiofu}. $\Phi(u)$ is obtained by integrating \eqref{graveqn1}. The integration constant $\Phi_0$ in the UV region which is simply $\Phi(u= \infty)$ can be determined to be about 6.91.

\begin{figure}[ht]
\centering
\includegraphics[width=0.8\textwidth]{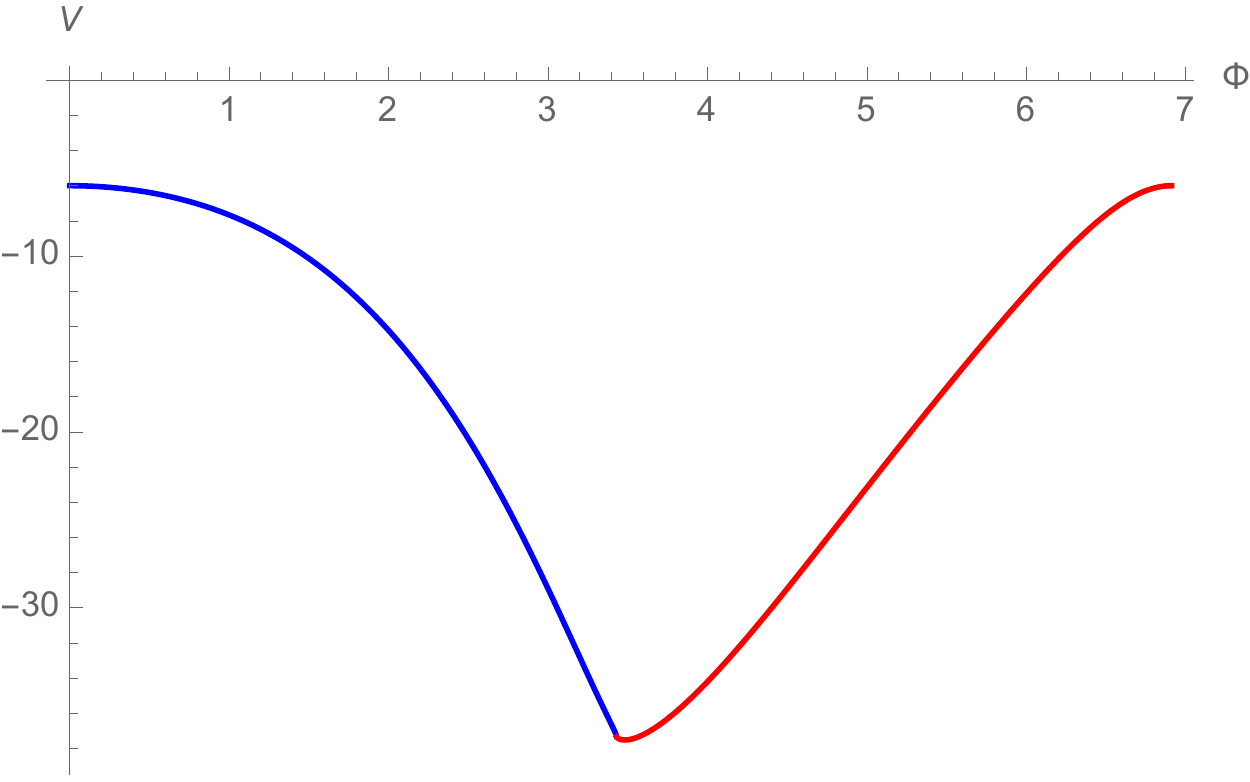}
\caption{\label{VPhi}Plot of $V(\Phi)$ (in units $L^{\rm UV} = 1$) as a function of $\Phi$; the blue curve refers to the UV region while the red one refers to the IR region. The kink in the middle where $V(\Phi)$ has a minima corresponds to $u=0$ where $\Phi = 3.43$ approximately. Here $V(\Phi)$ is not differentiable. The two asymptotic values of $V(\Phi)$ at the critical points $\Phi = 0$ and $\Phi \approx 6.91$ respectively are -6.}
\end{figure}

Finally, we plot $V(\Phi)$ in units $L^{\rm UV} =1$ as a function of $\Phi$ in Fig. \ref{VPhi}. We find that $V$ is $V-$shaped. The asymptotic values of $V(\Phi)$ in each component Universe is $-6$ where $V(\Phi)$ has critical points. The critical point in the UV universe is at $\Phi = 0$ and that in the IR universe is at $\Phi = \Phi_0 \approx 6.91$. Crucially, $V(\Phi)$ has a minima at $u= 0$ where $\Phi \approx 3.43$. Here $V(\Phi)$ is not differentiable, but still it is kind of a critical point as in the asymptotic regions. Furthermore, as clear from Figs. \ref{fig:rhopp} and \ref{fig:phiofu} that at $u=0$, $\rho'' = 0$ and $\Phi'= 0$ like in the two asymptotic regions also. Therefore, \textit{we can think of the region $u= 0$ as an AdS space of zero volume}. This hints that our full theory flows to an infrared fixed point. We leave a more detailed analysis to the future. 

If we take the perspective mentioned before that the UV (blue) and IR (red) universes are two covers of $-\infty < u \leq 0$ joined smoothly at $u = 0$, then clearly the two covers do not only have two different metrics but also two different potentials for the scalar field $\Phi$.

One final remark regarding determining all parameters of the IR theory and the hard-soft couplings as functions of the parameters of the UV theory is that we have assumed that the  irrelevant IR operator coupling to the UV operator has dimension $5$. Clearly, if we change the dimension of the IR operator to $6$ as for instance and modify our ansatz \eqref{rhoIR} accordingly, we will still be able to repeat the same exercise to obtain the new IR parameters and the hard-soft couplings. Thus the IR theory that completes the UV theory is unique up to certain assumptions of which the most crucial one is the scaling dimension of the irrelevant IR operator. In the case of semi-holographic framework for QCD, it will turn out that the dimensions of the IR operators to which the UV operator couples to will be fixed by perturbation theory itself. We will discuss this in the next section.
\paragraph{Additional comments:} Here we have assumed that both the UV and IR theories have holographic duals in the form of two-derivative gravity theories as both are sufficiently strongly coupled. It will be interesting to redo our construction after introducing higher derivative corrections to the dual (two-derivative) gravitational theories in both sectors. Firstly, the higher derivative corrections in the UV gravitational sector will determine those in the IR gravitational sector and also the hard-soft couplings through matching of bulk fields on the gluing hypersurface (analogous to $u= 0$ in our construction above) and our consistent coupling rules. Secondly, we speculate that higher derivative corrections in both sectors will make us interpolate both the effective UV and IR dynamics to weak coupling (as in usual string-theoretic examples of holography). In particular, in the weak coupling limit, bi-holography would then reduce to an explicit dual realisation of a form of Seiberg duality \cite{Intriligator:1995au} where both the electric (UV) theory and magnetic (IR) theory will be weakly coupled. This will also lead us to find explicit field-theoretic realisations of our bi-holographic constructions. We leave this investigation to the future.

\subsection{Excited states}\label{es}
As we have defined the bi-holographic theory and have explicitly constructed the vacuum state, we can proceed to compute physical observables of excited states. Let us first see how we can compute small fluctuations about the vacuum state. The parameters of the IR theory and the hard-soft couplings $\alpha$ and $\beta$ (at $\Lambda = \infty$) have been determined once and for all in terms of the parameters of  the UV theory. In fact these parameters together define the biholographic theory. Let us first consider scalar fluctuations, i.e. $\delta \rho^{\rm UV, IR}(u, x)$ and  $\delta \Phi^{\rm UV, IR}(u, x)$ in the UV and IR universes. As a result of the fluctuations, we generate $\delta\sigma^{(1)}(x)$, $\delta\sigma^{(2)}(x)$, $\delta T^{(1)}(x)$, $\delta O^{(1)}(x)$ and $\delta O^{(2)}(x)$. As discussed before, we should have $\delta J^{(2)}(x) = 0$, and the CFT Ward identity then implies that $\delta T^{(2)}(x) = 0$. In any case, we should solve the fluctuations so that perturbations of both sides of the semi-holographic coupling equations \eqref{eqn:sigma1n},  \eqref{eqn:sigma2n},  \eqref{eqn:J1n} and  \eqref{eqn:J2n} match. Crucially, we note that we are neither perturbing the fixed background metric $\eta_{\mu\nu}$ where the conserved energy-momentum tensor of the full system lives, nor adding any external source to the system. Individually in each Universe, we get two conditions each for each of the two sources (boundary metric and scalar source) from the coupling equations. The remaining conditions that we should impose will be that the perturbations must not affect the smooth gluing of the two Universes at $u= 0$. To this end, we will demand that at $u=0$
\be
\delta \rho^{\rm UV}(x) = \delta \rho^{\rm IR}(x), \quad {\delta \rho'}^{\rm UV}(x) = -{\delta \rho'}^{\rm IR}(x), \quad \text{and}\quad {\delta \Phi'}^{\rm UV}(x) = -{\delta \Phi'}^{\rm IR}(x).
\ee
Note that we have reversed the orientation of the radial direction in the IR universe before gluing as before.\footnote{A more diffeomorphism invariant statement is that the induced metrics and the Brown-York tensors on both sides should match at $u=0$ after we flip the sign of the Brown-York tensor on the IR side.} In order to ensure the continuity of $\Phi$ at $u=0$, we have to readjust the integration constant for $\Phi$ in the IR, i.e introduce an appropriate $\delta \Phi_0(x) \equiv \delta \Phi(u = \infty, x)$. This means that the potential $V^{\rm IR}(\Phi - \Phi_0 - \delta \Phi_0(x))$ is the same function as in the vacuum although the IR field $\Phi$ has now been redefined.\footnote{Note this canonical $V^{\rm IR}$ which remains state invariant is different from the red curve in Fig. \ref{VPhi}. In order to see this form we simply need to compute $V(\Phi -6.91)$ in the red region.} This redefinition means that $V^{\rm IR}$ has no tadpole term. However, this field-redefinition thus affect the definition of the IR theory in a subtle but concrete way. One can check that with these conditions, we can completely determine any fluctuation about the vacuum state and compute the perturbation of the full energy-momentum tensor of the dual system, etc.

It is easy to generalise the above discussion to the case of tensor and vector fluctuations of the vacuum state. Furthermore, we can similarly consider fluctuations of other bulk fields which vanish in the vacuum solution. In order to generalise our construction of  excited states which are not small departures from the vacuum, we can still use the general coupling conditions \eqref{eqn:sigma1n},  \eqref{eqn:sigma2n},  \eqref{eqn:J1n} and  \eqref{eqn:J2n}. However, we cannot use $u=0$ as the matching hypersurface as the domain-wall coordinates in which we have constructed the vacuum solution will be ill-defined beyond some patches of the UV and IR components individually. In the general case, \textit{we postulate that the UV and IR universes should be glued at their edge hypersurfaces where there is no curvature singularity but beyond which solutions for the matter fields cease to exist.} This gluing will then remove the edge singularities in each individual component which arises from geodetic incompleteness as in the vacuum case. In order for the postulate to make sense, \textit{we would require that edge singularities should appear in each component Universe much before any curvature singularity can occur.} Although the matter fields will not exist beyond the edge singularities, the individual UV and IR metrics can also be continued in the unphysical regions as we have seen in the case of the vacuum. Event horizons can lie either in the physical or in the unphysical parts of each component Universe system. At this stage, we are not sure what should be the general thermodynamic description of bi-holographic thermal states, although armed with our well-defined full energy-momentum tensor we can in principle study this question. It will be also fascinating to understand non-equilibrium behaviour of bi-holographic systems. We leave such investigations for the future.

We need to discuss though how we can couple external sources to the full bi-holographic system. Since we have already considered the case of consistent coupling rules when the fixed background metric is an arbitrary curved metric in the previous section, we need to understand only how to introduce other external scalar sources and external gauge potentials. We consider the case of external scalar sources only, as we have not studied the case of vector-type couplings. In presence of an external scalar source $J^{\rm ext}$, we need to modify the general coupling rules \eqref{eqn:sigma1n},  \eqref{eqn:sigma2n},  \eqref{eqn:J1n} and  \eqref{eqn:J2n} to:
\begin{subequations}
\label{allequations}
\begin{align}
e^{4\sigma^{(1)}} ~ &= ~ 1 +4 \beta \left(T^{(2)} + O^{(2)}\right) ,\label{eqn:sigma1nn}\\
e^{4\sigma^{(2)}} ~ &= ~ 1+ 4 \beta  \left(T^{(1)} + O^{(1)}\right), \label{eqn:sigma2nn}\\
J^{(1)} ~ &= ~ \frac{1}{4}{\rm ln} \left(1 +4\beta\left(T^{(2)} + O^{(2)}\right)\right) + \alpha O^{(2)} + J^{\rm ext},\label{eqn:J1nn}\\
0~ &= ~ \frac{1}{4}{\rm ln} \left(1 + 4\beta\left(T^{(1)} + O^{(1)}\right)\right) + \alpha O^{(1)} + J^{\rm ext}\, . \label{eqn:J2nn}
\end{align}
\end{subequations}
One can readily check that as a result of the above, the Ward identity of the full system will be modified to:
\begin{equation}
\partial_\mu T^\mu_{\phantom{\mu}\nu} = O \partial_\nu J^{\rm ext},
\end{equation}
where $T^\mu_{\phantom{\mu}\nu}$ will be given by the more general expression \eqref{eqn:totalscalartmunu} (with $g_{\mu\nu} = \eta_{\mu\nu}$ and $d= 4$) and 
\begin{equation}
O = O^{(1)} e^{4\sigma^{(1)}} + O^{(2)} e^{4\sigma^{(2)}}.
\end{equation}
In fact, this gives as a way to define the full operator $O$ of the biholographic (or semiholographic) system as a combination of the individual operators of the two sectors. More generally, $O$ will be\footnote{We can readily verify that in the most general case we need to add $J^{\rm ext}$ both to $J^{(1)}$ and $J^{(2)}$ in \eqref{tensor-scalar-couple} to obtain the general consistent coupling rules   in the presence of an external source $J^{\rm ext}$. It does not affect the general expression \eqref{eqn:totalscalar+tensortmunu} of the full energy-momentum tensor, however the Ward identity that it satisfies now should be $\nabla_\mu T^\mu_{\phantom{\mu}\nu} = O \nabla_\nu J^{\rm ext}$ in the fixed background metric $g$.  } 
\begin{equation}
O = O^{(1)} \sqrt{{\rm det }z^{(1)}} + O^{(2)} \sqrt{{\rm det }z^{(2)}},
\end{equation}
i.e. the sum of the individual operators weighted by the individual volume density factors of the effective metrics (recall $\sqrt{{\rm det }z^{(i)}} = \sqrt{{\rm det} g^{(i)}}/\sqrt{{\rm det}g}$). Thus $J^{\rm ext}$ couples democratically also to both sectors -- the relative strengths of the couplings being determined dynamically by the compression/dilation factors of the volume densities of the individual effective metrics as compared to the fixed background metric. With the coupling rules now set by \eqref{eqn:sigma1nn},  \eqref{eqn:sigma2nn},  \eqref{eqn:J1nn} and  \eqref{eqn:J2nn}, we can repeat the discussion before about how to compute small perturbations of the biholographic vacuum state and also other states far away from the vacuum.

\subsection{The highly efficient RG flow perspective}

A natural question to ask is how we can achieve a RG flow description of the biholographic theory. The right framework is indeed \textit{highly efficient RG flow} as introduced in \cite{Behr:2015yna,Behr:2015aat} (for a recent short review see \cite{Mukhopadhyay:2016fre}) which has been shown to reproduce the traditional holographic correspondence. In particular, this framework will allow us to define a conserved energy-momentum tensor of the full system at each scale without the need for introducing an action formalism. One of the key points of construction of highly efficient RG flow is that we should allow also the background metric $g_{\mu\nu}(\Lambda)$ and sources $J(\Lambda)$ evolve with the scale $\Lambda$ as a \textit{state-independent} functionals $g_{\mu\nu}[T^{\alpha\beta}(\Lambda), O(\Lambda), \Lambda]$ and $J[T^{\alpha\beta}(\Lambda), O(\Lambda), \Lambda]$ of the scale and effective operators so that at each scale $\Lambda$, the Ward identity
\begin{equation}
\nabla_{(\Lambda)\mu} T^\mu_{\phantom{\mu}\nu}(\Lambda) = O(\Lambda)\nabla_{(\Lambda)\nu} J(\Lambda)
\end{equation}
is satisfied in the effective background $g_{\mu\nu}(\Lambda)$. Thus the effective background metric preserves the Ward identity. Such a RG flow can be non-Wilsonian and an explicit construction can be achieved in the large $N$ limit by defining single-trace operators via collective variables (instead of the elementary quantum fields) which parametrise their expectation values in all states. A highly efficient RG flow leads to a $(d+1)-$dimensional spacetime with $g_{\mu\nu}(\Lambda)$ being essentially identified with $\Lambda^{-2}\gamma_{\mu\nu}$, with $\gamma_{\mu\nu}$ being the induced metric on the hypersurface $r =\Lambda^{-1}$ at a constant value of the radial coordinate that is identified with the inverse of the scale. Furthermore, the dual $(d+1)-$ dimensional metric will follow diffeomorphism invariant equations with a specific type of gauge fixing that can be decoded from a deformed form of Weyl invariance associated with the corresponding highly efficient RG flow.

In the bi-holographic case, the \textit{highly efficient RG flow} construction should work by choosing correlated hypersurfaces $\Sigma_1$ and $\Sigma_2$ in the UV and IR universes for each scale $\Lambda$ as shown in Fig. \ref{fig:UVandIRUniverses}. These hypersurfaces should be given by the two equations:
\begin{equation}
\Sigma_1 := u^{\rm UV} = u^{\rm UV}(\Lambda, x), \quad \Sigma_2 := u^{\rm IR} = u^{\rm IR}(\Lambda, x).
\end{equation}
Furthermore, we can invoke new hypersurface coordinates via diffeomorphisms defined on each hypersurface
\be\label{hyper}
{x'}^{\rm UV, IR} = {x'}^{\rm UV, IR}(\Lambda, x).
\ee
Choosing these $2(d+1)$ functions we may be able to define the two hypersurfaces $\Sigma_1$ and $\Sigma_2$ and also hypersurface coordinates in the UV and IR universes such that \textit{there exists a reference metric $\gamma_{\mu\nu}(\Lambda)$ at each $\Lambda$ with respect to which the induced metrics $\gamma^{(i)}_{\mu\nu}$ and $\gamma^{(2)}_{\mu\nu}$ on $\Sigma_1$ and $\Sigma_2$ respectively will be correlated with the general coupling rules \eqref{tensor-scalar-couple} so that the existence of a conserved energy-momentum tensor of the full system at each scale taking the form \eqref{eqn:totalscalar+tensortmunu} in the effective background $\gamma_{\mu\nu}(\Lambda)$ is ensured. The latter follows from the coupling rules because diffeomorphism invariance of the classical gravity equations in each Universe implies that the Brown-York stress tensors (renormalised by covariant counterterms) on each hypersurface is conserved in the background metrics  $\gamma^{(1)}_{\mu\nu}(\Lambda)$ and $\gamma^{(2)}_{\mu\nu}(\Lambda)$ respectively. }\footnote{Note that actually we also need to define a reference effective source $J^{\rm ext}(\Lambda)$ along with the reference metric $\gamma_{\mu\nu}(\Lambda)$ and consider modified coupling rules  \eqref{eqn:sigma1nn}-\eqref{eqn:J2nn} between $\gamma^{(i)}$s and $J^{(i)}$s on the two hypersurfaces \eqref{hyper}. We have kept this implicit in this discussion to avoid over-cluttering of words. Also note that $J^{\rm ext}(\Lambda)$ need not vanish at finite $\Lambda$ even when it vanishes at $\Lambda = \infty$. } 

This highly efficient RG flow construction is clearly possible only for $d \leq 4$ because otherwise with the $2(d+1)$ functions specifying the hypersurfaces and hypersurface coordinates we may not be able to solve for the right background metric $\gamma_{\mu\nu}(\Lambda)$ which has $d(d+1)/2$ independent components. Furthermore, \textit{the effective hard-soft coupling constants featuring in the general coupling rules \eqref{tensor-scalar-couple} should not only be scale but also be state-dependent except for the case $\Lambda = \infty$}. At $\Lambda = \infty$, the hypersurfaces $\Sigma_1$ and $\Sigma_2$ are the conformal boundaries of the UV and IR universes respectively. Here the hard-soft couplings remain same as in the vacuum state and indeed these are used to then construct all excited states of the theory as mentioned above. Furthermore, at $\Lambda = \infty$, $\gamma_{\mu\nu}(\Lambda)$ simply coincides with the background metric on which the full system and it's energy-momentum tensor lives by construction. It is not clear if such a RG flow perspective makes sense for $d>4$, i.e. in bi-AdS spaces with more than 5 dimensions.

The highly efficient RG flow perspective gives a very coherent view of the full biholographic construction. In particular, by construction it breaks the apparent independent $(d+1)-$diffeomorphism invariance of the two Universes into only one kind of $(d+1)-$diffeomorphism invariance. The invariance of the conservation equation for the full energy-momentum in a reference metric background gives $d-$constraints. An additional Hamiltonian constraint arises naturally in order to form a first class constraint system. These $(d+1)-$constraints result in having $(d+1)-$diffeomorphism symmetry instead of twice the number.

One more attractive feature of the highly efficient RG flow construction is that one can take the point of view that spacetime emerges from the endpoint of the RG flow corresponding to the horizon of the emergent geometry rather than from the boundary. Imposing that the end point of the RG flow under an universal rescaling of the scale and time coordinate (corresponding to zooming in the long time and near horizon limits of the dual spacetime) can be mapped to a fixed point with a few parameters, we obtain bounds for the first order flows of effective physical observables near the end point such that at the boundary they take the necessary physical values which ensures absence of naked singularities in the dual spacetime \cite{Kuperstein:2013hqa,Behr:2015yna,Behr:2015aat}. This has been explicitly demonstrated in the context of the hydrodynamic limit of the dynamics in the dual quantum system specially. Taking such a point of view is natural in the bi-holographic context, because it is only at the matching hypersurface ($u=0$) of the two Universes corresponding to the endpoint of the highly efficient RG flow the two Universes physically overlap and share common data. Therefore, the two Universes naturally emerge from the $u=0$ hypersurface. In the future, we will like to investigate the RG flow reconstruction of bi-holography and also investigate if one can define $c-$functions for such RG flows.

\section{Concluding remarks}\label{sec:outlook}
\subsection{How to proceed in the case of QCD?}
The bi-holographic construction provides an illuminating illustration of how the semi-holographic framework can be derived from first principles, particularly regarding how some simple consistency rules can be used to determine the parameters of the IR holographic theory and the hard-soft couplings in terms of the parameters of the UV theory. Let us discuss briefly how the steps of the construction of bi-holography can be generalised  to the case of the semi-holographic framework for QCD.

Firstly, in the case of bi-holography the IR Universe was necessary to cure the edge singularity of the UV Universe, and the smoothness of the gluing between the two Universes was a key principle that determined the parameters of the gravitational theory of the IR Universe as well. In the case of QCD, an analogous issue is the cure the non-Borel resummability of the perturbation series, i.e. we need the non-perturbative (holographic) physics to cancel the \textit{renormalon} Borel poles of perturbation theory that lie on the positive real axis and control large order behaviour of the perturbation series in the large $N$ limit \cite{Beneke:1998ui,citeulike:9712091}. It is known that each such renormalon pole that appears in the perturbative calculations of the operator product expansion (OPE) of a product of two gauge-invariant operators can be cancelled by invoking a non-perturbative condensate of an appropriate gauge-invariant operator with the right mass dimension \cite{Parisi:1978bj,citeulike:9712091} as originally observed by Parisi. The further the renormalon pole is from the origin, the larger the mass dimension of the operator whose condensate cancels this pole should be. Furthermore, \textit{the non-perturbative dependence of this condensate on $\Lambda_{\rm QCD}$, or equivalently on the perturbative strong coupling constant is completely determined by pQCD (in particular by the location of the corresponding Borel pole that gets cancelled).}  

This observation of Parisi can be transformed into a physical mechanism via semi-holography. In particular the non-perturbative condensate of a given operator with  given mass dimension should be reproduced by the dynamics of the dual holographic bulk field. Since the condensate is determined by perturbation theory, the holographic gravitational theory should be \textit{designed} appropriately in order to reproduce the right behaviour as a function of the confinement scale. Furthermore, the gravitational boundary condition determined by the hard-soft coupling(s) with the corresponding operator of the perturbative sector appearing in the perturbative expansion of the OPE must also be specified in an appropriate way. Such a \textit{designer gravity} approach for designing a holographic gravitational theory and its boundary conditions in order to reproduce right behaviour of the dual condensates has been studied in \cite{Hertog:2004ns,Gubser:2008yx,Gursoy:2008za}.  This approach can be adapted to the semi-holographic construction to determine the holographic theory dual to the non-perturbative sector, and also the hard-soft couplings between the perturbative and the non-perturbative sectors as functions of $\Lambda_{\rm QCD}$.

It is clear that the construction of this semi-holographic framework should be far more complicated than in the bi-holographic case. Multiple number of non-perturbative condensates, i.e. operators of the emergent holographic theory should couple to each gauge-invariant operator of the perturbative sector. However, we can proceed systematically by considering the cancellation of perturbative Borel poles in closer proximity to the origin for which we would require non-perturbative condensates of lower mass dimensions only.

The Borel poles of the (appropriately resummed) perturbation theory can shift at scales intrinsic to a non-trivial state (as for instance the temperature). This naturally implies that we need to invoke state-dependence in the running of the hard-soft couplings with the scale. The bi-holographic construction further indicates that we need to do field-redefinitions in the holographic gravitational theory which could be state-dependent although its parameters should be not vary with the state. We need to understand these issues better in the future. However, the arguments presented above indicate that the semi-holographic framework for QCD can indeed be constructed systematically in the large $N$ limit.

Physically, the gravitational sector should naturally give rise to the effective low energy degrees of freedom of QCD, namely the mesons, baryons and glueballs as in case of the top-down Witten-Sakai-Sugimoto model \cite{Witten:1998zw,Sakai:2004cn,Sakai:2005yt,Brunner:2015oqa} and other holographic QCD models \cite{Erlich:2005qh}. The eigen-modes of  linearised perturbations of the holographic gravitational \textit{vacuum} sector can be identified with the observed mesons, baryons and glueballs based on their quantum numbers. Furthermore, the holographic gravitational sector naturally produces a geometric realisation of confinement \cite{Witten:1998zw,Karch:2006pv} which cannot be obtained easily from Feynman diagrams.

It is clear that since we are proposing to construct the holographic gravitational dual of QCD from first principles, we cannot assume any form of asymptotic behaviour of the emergent spacetime describing the non-perturbative sector. The latter should be a consequence of the gravitational equations which should be \textit{designed} to cancel the Borel singularities of the QCD perturbation series. Since QCD is not a conformal theory, the gravitational spacetime is unlikely to be asymptotically anti-de Sitter. However, the requirement of having a well-defined holographic renormalisation procedure for defining the vacuum condensates of gauge-invariant operators implies that the scale factor of the geometry should have appropriate singularities in the asymptotic limit. Such a form of modified asymptotic structure is indeed implemented in some bottom-up holographic QCD approaches eg. improved holographic QCD \cite{Gursoy:2008za}.

\subsection{Possible applications of bi-holography}
In the present paper, we have invoked bi-holography to illustrate semi-holography. However, it should be worthwhile to pursue the bi-holographic framework and its applications on its own right. In particular, the bi-holographic framework gives rise to a consistent bi-metric gravitational theory as mentioned before. Such a construction when invoked in the context of positive (instead of negative) cosmological constant, can perhaps also be relevant for shedding light on the origin of dark matter (in the form of matter in the ghost Universe which gives the second covering of the full spacetime). It has also been pointed out in the literature that the possibility that baryonic matter and dark matter can live in different effective metrics can explain late time acceleration of the Universe without invoking the cosmological constant \cite{Berezhiani:2016dne}. The visible Universe with baryonic matter, and the coexisting ghost Universe with dark matter and the second metric should be joined at the beginning of time such that each can cure the other's initial-time singularity. It might be interesting to pursue such a cosmological model.

Finally, bi-holography can have applications which are more wide ranging than holography. In particular the effective metrics on which the UV and IR sectors live can have different topologies from the original background metric\footnote{This is specially relevant when the tensorial hard-soft couplings $\gamma_1 \neq 0$ and/or $\gamma_2 \neq 0$. In non-relativistic bi-holography, we will not be able to set these to zero even in the limit $\Lambda\rightarrow\infty$ in order to construct the vacuum state.} which should be determined dynamically. This can then serve as examples of theories with hidden topological phases which cannot be captured by local order parameters, and admitting simple geometric descriptions. With such applications in view, it should be interesting to study bi-holographic RG flows, and also thermal and non-equilibrium dynamics in bi-holography.

We conclude with the final remark that we must also pursue if bi-holography and semi-holography can be embedded in the string theoretic framework. This direction of research may extend the horizons of string theory, and may also lead to a deeper and more enriching understanding of the field-theoretic implications of bi-holography and semi-holography.

\begin{acknowledgments}
N.G would like to thank Panos Betzios, Umut Gursoy, Olga Papadoulaki and Tatjana Pu\v{s}karov for several helpful conversations. The research of N.G. is supported by Netherlands Organisation for Scientific Research (NWO) under the VICI grant 680-47-603, and the Delta-Institute for Theoretical Physics (D-ITP) that is funded by the Dutch Ministry of Education, Culture and Science (OCW). S.B would like to thank Giuseppe Dibitetto and Ioannis Papadimitriou for helpful conversations. The research of S.B. is supported by the Knut and Alice Wallenberg Foundation under grant 113410212. The research of A.M. is supported by a Lise-Meitner fellowship of the Austrian Science Fund (FWF), project no. M 1893-N27. A.M. thanks Jorge Casalderrey-Solana, Jerome Gauntlett, Edmond Iancu, Elias Kiritsis, David Mateos, Vasilis Niarchos, Marios Petropoulos, Giuseppe Policastro, Florian Preis and Anton Rebhan for exchange of ideas and helpful discussions. We thank Florian Preis and Anton Rebhan for comments on the manuscript. We also thank Kaushik Bandyopadhyay for helping us with the artwork of Fig. \ref{fig:UVandIRUniverses}.
\end{acknowledgments}

\bibliographystyle{apsrev4-1}
\bibliography{PrinciplesPaper_SB}
\end{document}